\documentclass[preprint,3p,times]{elsarticle}

\usepackage{amssymb}
\usepackage{amsthm}
\usepackage{amsmath}
\usepackage{epsfig}
\usepackage{subfigure}
\usepackage[usenames]{color}
\usepackage{algorithm}
\usepackage{algorithmic}
\usepackage{nicefrac}
\usepackage{caption}
\epsfverbosetrue
\graphicspath{{fig/}{../fig/}}

\usepackage[]{natbib}

\newcommand{\mb}[1]{\ensuremath{\mathbf{#1}}}

\journal{Journal of Materials Processing Technology}

\begin{document}
 \setlength{\parindent}{0.0ex}
 \setcounter{secnumdepth}{4}
 \setcounter{tocdepth}{4}
\begin{frontmatter}






\title{Critical Influences of Particle Size and Adhesion on the Powder Layer Uniformity in Metal Additive Manufacturing}




\author[mechanosynth,lnm]{Christoph Meier\corref{cor1}}
\ead{meier@lnm.mw.tum.de}
\author[mechanosynth,lnm]{Reimar Weissbach}
\author[mechanosynth,lincoln]{Johannes Weinberg}
\author[lnm]{Wolfgang A. Wall}
\author[mechanosynth]{A. John Hart\corref{cor1}}
\ead{ajhart@mit.edu}

\address[mechanosynth]{Mechanosynthesis Group, Department of Mechanical Engineering, Massachusetts Institute of Technology, 77 Massachusetts Avenue, Cambridge, 02139, MA, USA}
\address[lnm]{Institute for Computational Mechanics, Technical University of Munich, Boltzmannstrasse 15, 85748 Garching b. M{\"u}nchen, Germany}
\address[lincoln]{M.I.T. Lincoln Laboratory, 244 Wood Street, Lexington, 02420, MA, USA}

\cortext[cor1]{Corresponding authors}


\bibliographystyle{recoating1b.bst}
\setcitestyle{authoryear}

\begin{abstract}
The quality of powder layers, specifically their packing density and surface uniformity, is a critical factor influencing the quality of components produced by powder bed metal additive manufacturing (AM) processes, including selective laser melting, electron beam melting and binder jetting. The present work employs a computational model to study the critical influence of powder cohesiveness on the powder recoating process in AM. The model is based on the discrete element method (DEM) with particle-to-particle and particle-to-wall interactions involving frictional contact, rolling resistance and cohesive forces. Quantitative metrics, namely the spatial mean values and standard deviations of the packing fraction and surface profile field, are defined in order to evaluate powder layer quality. Based on these metrics, the size-dependent behavior of exemplary plasma-atomized Ti-6Al-4V powders during the recoating process is studied. It is found that decreased particle size / increased cohesiveness leads to considerably decreased powder layer quality in terms of low, strongly varying packing fractions and highly non-uniform surface profiles. For relatively fine-grained powders (mean particle diameter $17 \mu m$), it is shown that cohesive forces dominate gravity forces by two orders of magnitude leading to low quality powder layers not suitable for subsequent laser melting without additional layer / surface finishing steps. Besides particle-to-particle adhesion, this contribution quantifies the influence of mechanical bulk powder material parameters, nominal layer thickness, blade velocity as well as particle-to-wall adhesion. Finally, the implications of the resulting powder layer characteristics on the subsequent melting process are discussed and practical recommendations are given for the choice of powder recoating process parameters. While the present study focuses on a rigid blade recoating mechanism, the proposed simulation framework can be applied to and the general results can be transferred to systems based on alternative recoating tools such as soft blades, rakes or rotating rollers.
\end{abstract}
\end{frontmatter}

%
%
\section{Introduction}
\label{sec:intro}
%
%

Among the manifold of existing additive manufacturing (AM) processes, selective laser melting (SLM) of metals has attracted much scientific attention because it offers near-net-shape production of near-limitless geometries, and eventual potential for pointwise control of microstructure and mechanical properties~\citep{Gibson2010}. However, the overall SLM process is complex and governed by a variety of (competing) physical mechanisms. A sub-optimal choice of process parameters might lead to deteriorated material properties or even to failure of the part already during the manufacturing process~\citep{Das2003,Kruth2007}. For that reason, a multitude of experimental and modeling approaches have been conducted in recent years in order to gain further understanding of the underlying physical mechanisms and ultimately to optimize the process and the properties of the final part.\\

Depending on the considered lengths scales, existing experimental and modeling approaches can typically be classified in the three categories macroscopic, mesoscopic and microscopic~\citep{Meier2018}. Macroscopic approaches commonly aim at determining spatial distributions of physical fields such as temperature, residual stresses or dimensional warping on part level. In one of the pioneering works in this field,~\cite{Gusarov2007} solved the thermal problem of the SLM process based on a previously developed radiation transfer model for laser energy absorption~\citep{Gusarov2005}. \cite{Hodge2014} extended this procedure to a coupled thermo-(solid-)mechanical analysis employing a (homogenized) elasto-plastic constitutive law accounting for thermal expansion and consolidation shrinkage. \cite{Denlinger2014} supplemented the contributions discussed so far by allowing for residual stress relaxation. In order to improve computational efficiency, \cite{Riedlbauer2016} applied adaptive mesh refinement strategies, while \cite{Zaeh2010} considered a simplified process model based on equivalent thermal loads applied to several powder layers at once. Mesoscopic approaches commonly predict resulting melt pool and part properties such as melt track stability, surface quality or layer-to-layer adhesion as well as creation mechanism of defects such as pores and inclusions having their origin on the mesoscopic scale of individual powder particles. \cite{Khairallah2014} proposed a model of this type based on the finite element method (FEM) considering viscous, gravity and surface tension forces, which has been extended by \cite{Khairallah2016} to also account for the effects of recoil pressure in the evaporation zone below the laser beam, Marangoni convection, as well as evaporative and radiative surface cooling. There is a range of alternative discretization schemes such as the finite difference method (FDM), see e.g.~\cite{Lee2015}, the finite volume method (FVM), see e.g.~\cite{Qiu2015}, or the Lattice-Boltzmann method (LBM), see e.g. \cite{Korner2011}, that have been utilized in the context of mesoscopic models for the SLM process. Last, microscopic approaches consider the evolution of the metallurgical microstructure involving the resulting grain sizes, shapes and orientations as well as the development of thermodynamically stable or unstable phases. In this context, \cite{Gong2015} applied a phase field model in order to study the solidification and growth of primary $\beta$-phase grains during the processing of Ti-6Al-4V. As alternative to the mentioned phase field model, \cite{Rai2016} proposed a microscale model based on a cellular automaton (CA) scheme. The majority of existing approaches on the macro-, meso-, and microscale relates the mentioned process outcomes (e.g. residual stresses, surface quality, metallurgical microstructure etc.) with typical input parameters such as laser beam power and velocity, powder layer thickness, hatch spacing or scanning strategy, typically with the final goal to derive process maps~\citep{Thomas2016}.\\

Very recently, the importance of the primary input parameter powder feedstock has been underlined by political, scientific and industrial leaders in the field of AM technologies~\citep{King2017}. In fact, there are several works that have studied (the influence of) the powder feedstock, which is characterized by mechanical, thermal, optical and chemical properties on the surface of individual powder particles, by morphology, granulometry and resulting flowability of bulk powder, and, eventually, by the resulting packing density, surface uniformity and effective thermal and mechanical properties of the deposited powder layer. \cite{Herbert2016} gives an overview of important metallurgical aspects in the different stages of powder handling / treatment during the SLM process, i.e. in the sequence from powder storage, to spreading in the machines, to melting, solidification, and post-processing. \cite{Tan2017} extends this overview to more general aspects with a special focus on the influence of powder morphology and granulmetry. In the context of powder feedstock modeling,~\cite{Gusarov2008} studied the problem of laser energy absorption in powder beds based on a (homogenized) continuum model for the radiation transfer problem while~\cite{Boley2015} approached the same problem on the basis of a ray tracing scheme and a (discrete) powder bed model resolving individual particles. Unfortunately, only very few contributions have studied the interplay of the aforementioned powder particle, bulk and layer properties during the powder recoating process in metal additive manufacturing (see e.g. the experimental study of particle size segregation by~\cite{Whiting2016}). More specifically, very few modeling and simulation approaches of the powder recoating process in metal additive manufacturing can be found in the literature. For example,~\cite{Herbold2015} demonstrated that DEM simulations can reproduce realistic powder beds in terms of size-segregation and packing distributions and modeled different practically relevant blade geometries. In the recent work by~\cite{Mindt2016}, the influence of the blade gap on the resulting powder layer has been investigated. Their work has shown that small blade gaps in the range of the maximal powder particle diameter or below might lead to a considerably decreased packing density.~\cite{Haeri2017} investigated the influence of different recoating device geometries. While the studies were based on polymer materials, a transferability of general results to metal powders was claimed by the author of this study. Also~\cite{Gunasegaram2017} presented a general framework for the simulation of powder recoating / raking processes in metal AM. The mentioned models are based on the discrete element method (DEM) and typically account for (visco-) elastic normal contact, sliding friction as well as rolling friction interaction between spherical particles. Even though its importance has already been discussed by~\cite{Herbold2015}, cohesive effects within the powder have not been considered in any of these pioneering simulation works.\\

Considering that volume forces such as gravity decrease cubically with particle size while typical adhesive forces decrease linearly with particle size confirms the well-known observation that the cohesiveness of bulk powder increases with decreasing particle size~\citep{Walton2008}. Typical (adhesive) surface energy values measured for metals as well as typical powder particle size distributions applied in metal additive manufacturing suggest that adhesion might be an important factor governing the flowability of bulk powder~\citep{Herbold2015}. There are indeed modeling approaches, where cohesive effects have successfully been taken into account in DEM-based powder recoating studies, however, to the best of the authors' knowledge, only for non-metallic materials so far. For example,~\cite{Parteli2016} studied the spreading of non-spherical, polymeric particles and confirmed the well-known, yet counter-intuitive, effect that powder mixtures containing smaller particle species might result in lower effective packing densities due to cohesion-induced particle clustering~\citep{Walton2007}. Since the underlying physical mechanisms as well as the magnitude of cohesive effects (and also of other material properties) are crucially different for (conductive) metallic powders and non-metallic powders, it is questionable if results gained for the latter class of materials can directly be transferred to powders relevant for metal additive manufacturing.\\

Since surface energies of a given material combination can easily vary by orders of magnitude as consequence of surface roughness and potential surface contamination / oxidation, it is essential to calibrate the employed model by experimentally characterizing the considered powders in terms of effective surface energy values. In the authors' recent contribution,~\cite{Meier2018a} determined for the first time an effective surface energy value for the considered class of metal powders, which will be employed in the present study. Moreover, also the importance of consistently considering adhesive particle interactions has been demonstrated by~\cite{Meier2018a} by comparing the bulk powder behavior predicted by models with and without adhesive force contributions. This important result motivates the present work, representing the first numerical study of powder recoating processes in metal additive manufacturing based on a cohesive powder model. Thereto, proper metrics, namely the spatial packing fraction and surface profile field will be defined, in order to analyze the characteristics of powder layers in terms of mean values and standard deviations in these metrics. The focus of the present studies lies on the influence of powder cohesiveness, which essentially depends on the magnitudes of surface energy and mean particle size within the considered plasma-atomized Ti-6Al-4V powder. Consequently, the powder recoating process will be analyzed for three different powder size classes, i.e. three different classes of powder cohesiveness. Out of these three powder size classes, the most fine-grained powder with a mean particle diameter of $17 \mu m$ is characterized by cohesive forces that dominate gravity forces by two orders of magnitude. It is demonstrated that the resulting powder layers are of low quality in terms of the previously defined metrics and not suitable for subsequent laser melting without additional layer / surface finishing steps. Moreover, the influence of mechanical bulk powder material parameters, nominal layer thickness, blade velocity as well as particle-to-wall adhesion is quantified, and the interplay of these parameters with bulk powder cohesion is analyzed. Eventually, possible implications of the resulting powder layer characteristics on the subsequent melting process are briefly discussed and practical recommendations are given for the choice of powder recoating process parameters. These discussions are carried on with having the SLM process in mind, but the presented results are also valid for other representatives of metal-fusion based additive manufacturing processes, and might, in general, have relevance for all kinds of recoating and packing processes of micron-scale powders.\\

The remainder of this article is structured as follows: Section~\ref{sec:model} briefly recapitulates the essentials of the cohesive powder model proposed by~\cite{Meier2018a} and the choice of model parameters. In Section~\ref{sec:recoating}, this model is employed in order to analyze the influence of powder cohesiveness and other powder material and process parameters on the resulting powder layer quality. Finally, these results and implications on the subsequent melting process are discussed in Section~\ref{sec:discussion}, before a summary of the present contribution and a brief outlook on future research work is given in Section~\ref{sec:conclusion}.\\

\section{Cohesive powder model and choice of parameters}
\label{sec:model}

\begin{figure}[h!!]
 \centering
   {
    \includegraphics[height=0.33\textwidth]{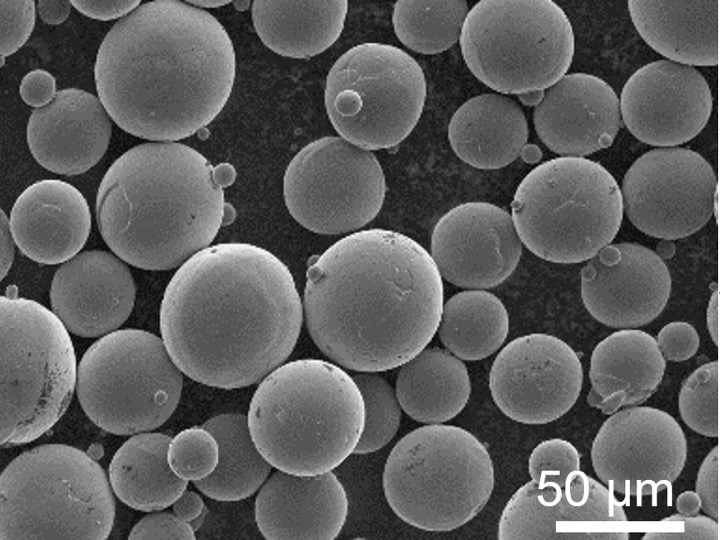}
    \label{fig:powder_SEM1}
   }
\hspace{0.5cm}
   {
    \includegraphics[height=0.33\textwidth]{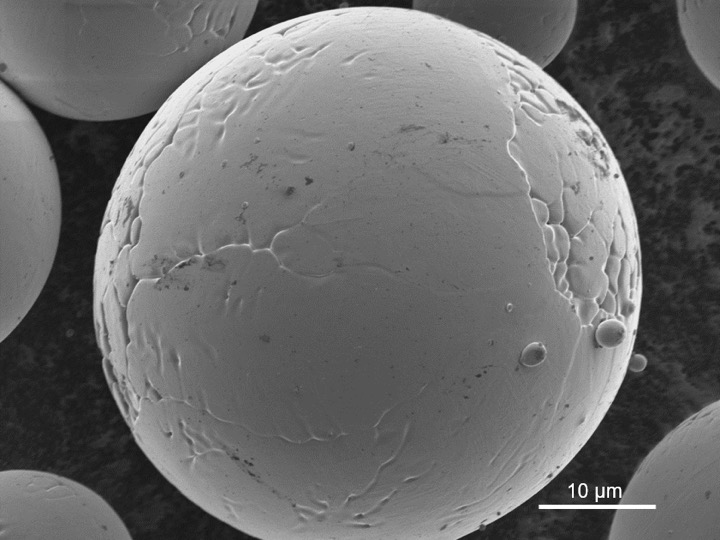}
    \label{fig:powder_SEM2}
   }
  \caption{SEM Images of employed plasma-atomized Ti-6Al-4V powder as typical for metal additive manufacturing processes,~\cite{Meier2018a}.}
  \label{fig:powder_SEM}
\end{figure}

In the recoating simulations conducted throughout this work, plasma-atomized Ti-6Al-4V powders with different size distributions are considered. According to the SEM images in Figure~\ref{fig:powder_SEM}, individual powder particles can be described as spherical particles in good approximation. The cohesive powder model proposed by~\cite{Meier2018a} relies on the discrete element method (DEM), going back to~\cite{Cundall1979}, more precisely on the so-called "soft sphere" model, in order to model the bulk powder on the level of individual powder particles in a Lagrangian manner. In general, the equations of motion of an individual particle are given by the balance of linear and angular momentum:
\begin{subequations}
\label{momentum}
\begin{align}
(m \, \ddot{\mb{r}}_G)^i = m^i \mb{g} + \sum_j (\mb{f}_{CN}^{ij}+\mb{f}_{CT}^{ij}+\mb{f}_{AN}^{ij}),\label{gusarov2007_HCE1}\\
(I_G \, \dot{\boldsymbol{\omega}})^i = \sum_j (\mb{m}_{R}^{ij}+\mb{r}_{CG}^{ij} \times \mb{f}_{CT}^{ij}).\label{gusarov2007_HCE2}
\end{align}
\end{subequations}
Here, $m\!=\!4/3\pi r^3\rho$ is the particle mass, $I_G=0.4mr^2$ is the moment of inertia of mass with respect to the particle centroid $G$, $r$ is the particle radius, $\rho$ is the mass density, $\mb{r}_G$ is the particle centroid position vector, $\mb{g}$ is the gravitational acceleration, and $\boldsymbol{\omega}$ represents the angular velocity vector. The notion $\dot{(...)}$ represents the first time derivative. Moreover, a velocity verlet time integration scheme is applied to discretize equations~\eqref{momentum} in time. The right-hand side of~\eqref{momentum} summarizes the forces and torques resulting from an interaction with neighboring particles $j$ considering normal and tangential contact forces $\mb{f}_{CN}^{ij}$ and $\mb{f}_{CT}^{ij}$, adhesive forces $\mb{f}_{AN}^{ij}$, rolling resistance torques $\mb{m}_{R}^{ij}$ as well as the torques resulting from the tangential contact forces. In this context, $\mb{r}_{CG}^{ij}:=\mb{r}_{C}^{ij}-\mb{r}_{G}^i$ represents the vector from the centroid of particle $i$ to the point of contact with particle $j$. For the specific force laws underlying these interactions and the choice of model parameters, the interested reader is referred to~\cite{Meier2018a}. Only the modeling parameters most relevant for the present study, will briefly be repeated in the following.\\

The time step size $\Delta t$ of the employed (explicit) velocity verlet time integration scheme has been chosen according to a critical step size estimate given by the following relation~\citep{OSullivan2004}
\begin{align}
\label{stepsize}
     \Delta t \leq \Delta t_{crit}= 0.2 \sqrt{\frac{m_{min}}{k_N}},
\end{align}
with $m_{min}:=\min_{i}(m)$ representing the minimal mass across all particles. While a finite element discretization has been chosen to represent rigid walls, the modelled particle size distribution is considered to be of log-normal type with parameters fitted to the material certificate of the employed powders. Specifically, the particle diameter distribution of the reference powder considered throughout this work is specified by $10\%$, $50\%$ and $90\%$ percentiles according to  $D10=20 \mu m$,  $D50=34 \mu m$ and $D90=44 \mu m$, representing a typical medium-sized powder for AM applications with mean particle diameter $\bar{d}=34\mu m$. After fitting the employed log-normal distribution to these specifications, the range of particles sizes considered in the model has been limited to lie within $D10$ and $D90$, i.e. very small particles below $d_{min}=20 \mu m$ as well as very large particles above $d_{max}=44 \mu m$ have not been considered for reasons of computational efficiency. The density of the Ti-6Al-4V particles is $\rho=4430 kg/m^2$, and the Hamaker constant required for the adhesion force law has been taken as $A=40\cdot 10^{-20} J$~\citep{Israelachvili2011}.\\

\cite{Meier2018a} calibrated the surface energy value required to calculate the adhesive interaction forces of the employed model for the considered Ti-6Al-4V powder by fitting experimental and numerical AOR values, leading to $\gamma_0=0.1mJ/m^2$. In order to analyze the influence of adhesion on the actual powder recoating process, the simulation of this process will not only be carried out for the parameter choice $\gamma=\gamma_0$, but also for $\gamma=4\gamma_0$, $\gamma=0.25\gamma_0$ and $\gamma=0$. The stiffness/penalty parameter $k_N$ of the contact model has been chosen such that the expected maximal relative penetrations $c_g$ between contacting particles lie below $2.5\%$~\citep{Meier2018a}. Concretely, a value of $k_N = 0.05N/m$ has been applied in all simulations with surface energies $\gamma \leq \gamma_0$, whereas the increased value $k_N = 0.2N/m$ has been employed in all simulations with surface energy $\gamma=4 \gamma_0$. Moreover, the friction coefficient and the coefficient of restitution for particle-to-particle and particle-to-wall interaction have been chosen as $\mu=0.4$ and $c_{COR}=0.4$ in the present study. Throughout this work, the notation particle-to-wall interaction includes all interactions of particles with surrounding solid components and walls. Specifically, it includes the interaction between particles and recoating blade (particle-to-blade interaction) as well as the interaction between particles and the bottom / side walls of the powder bed (particle-to-substrate interaction). Since only the surface energy value of the employed model has been calibrated on the basis of bulk powder experiments by~\cite{Meier2018a}, verification simulations based on varied parameter values will be conducted in Section~\ref{sec:recoating_variation} in order to assess the sensitivity of the resulting powder layer characteristics with respect to the aforementioned parameter choice. In accordance to the results by~\cite{Meier2018a}, it will turn out that the variation of these parameters (within a reasonable range) will typically have considerably lower influence on the resulting powder layer characteristics than variations of the surface energy, which lies in the focus of this work.\\

All simulation results presented in this work rely on a software implementation of the proposed DEM formulation in the in-house code BACI, a parallel multiphysics research code with finite element, particle and mesh-free functionalities, developed at the Institute for Computational Mechanics of the Technical University of Munich~\citep{Wall2018}.\\

\section{Recoating simulations}
\label{sec:recoating}

In this section, the powder recoating process in powder-fusion based metal additive manufacturing will be simulated. A special focus will lie on the influence of adhesion on the spreadability of the metal powder and the resulting powder layer characteristics. Besides the actual surface energy $\gamma=\gamma_0=0.1mJ/m^2$ of the considered medium-size powder (mean particle diameter $\bar{d}=34\mu m=:d_0$) as determined by~\cite{Meier2018a}, also the variations $\gamma=\gamma_0/4$ and $\gamma=4\gamma_0$ will be investigated. As shown by~\cite{Yang2000}, the characteristics (e.g. packing fraction, coordination number, surface uniformity) a layer of cohesive powder takes on in static equilibrium, can in good approximation be formulated as function of the dimensionless ratio of adhesive (pull-off) and gravity force, which scales quadratically with the particle diameter according to $F_{\gamma}/F_G \sim \gamma / (\rho g d^2) $. Thus, the increase / decrease of surface energy $\gamma$ by a factor of $4$ is equivalent to the decrease / increase of particle size $d$ by a factor of $2$, which allows to evaluate the influence of powder particle size on powder bed quality (see Section~\ref{sec:recoating}). Specifically, the variations $\gamma=\gamma_0/4$ and $\gamma=4\gamma_0$ are equivalent to increasing / decreasing the mean particle diameter from $\bar{d}=d_0=34\mu m$ to the values $\bar{d}=2d_0=68 \mu m$ and $\bar{d}=d_0/2=17 \mu m$, respectively. \cite{Meier2018a} confirmed this equivalence by comparing funnel simulations of the variants $\gamma=\gamma_0, \, \gamma=\gamma_0/4$ and $\gamma=4\gamma_0$ (mean particle diameter $\bar{d}=d_0=34\mu m$ in all simulations) with the experimental bulk powder behavior of the considered medium-sized Ti-6Al-4V powder ($\bar{d}=d_0=34\mu m$) as well as of a more coarse-grained and a more fine-grained powder fulfilling the scalings $\bar{d}=2d_0$ and $\bar{d}=d_0/2$ in good approximation (identical surface energy $\gamma=\gamma_0$ assumed for all three powders employed in the experiments).\\

In addition to the three cases $\gamma=\gamma_0, \, \gamma=\gamma_0/4$ and $\gamma=4\gamma_0$, also the (theoretical) case of vanishing adhesion $\gamma=0$, i.e. the model with no adhesion, will be investigated. As discussed above, the equivalence of increased / decreased adhesion and decreased / increased particle size can be motivated by the dimensionless adhesion-to-gravity force ratio $F_{\gamma}/F_G$, which is identical in both cases. For clarity, the force ratios considered in this section (calculated for the scenario of two contacting particles with diameter $\bar{d}$) as well as the associated equivalent surface energies and mean particle diameters are plotted in Table~\ref{tab:equivalent_gamma_vs_radius}. Again, the variation of surface energy at constant particle size $\bar{d}=d_0$ (first line of Table ~\ref{tab:equivalent_gamma_vs_radius}) represents the simulation strategy employed in the present work, whereas the variation of mean particle size at constant surface energy $\gamma = \gamma_0$ (second line of Table ~\ref{tab:equivalent_gamma_vs_radius}) rather equals a typical experimental approach.\\

\begin{table}[h!]
\centering
\begin{tabular}{|p{3.0cm}|p{1.5cm}|p{1.5cm}|p{1.5cm}|p{1.5cm}|} \hline
 $\bar{d}=d_0, \quad \quad \quad  \gamma / \gamma_0:$ & $0$ & $0.25$ & $1$ & $4$ \\ \hline
$\gamma = \gamma_0, \quad \quad \quad  \bar{d} / d_0:$ & $\infty$ & $2$ & $1$ & $0.5$ \\ \hline
$\quad \quad \quad  \quad \, \quad F_{\gamma}/F_G:$ & $0$ & $3.25$ & $13$ & $52$ \\ \hline
\end{tabular}
\caption{Surface energies and equivalent mean particle diameter and adhesion-to-gravity force ratios.}
\label{tab:equivalent_gamma_vs_radius}
\end{table}

From Table~\ref{tab:equivalent_gamma_vs_radius}, it becomes obvious that the cases $\gamma = \gamma_0$ and $\gamma = 4 \gamma_0$ are characterized by adhesion forces that are one to two orders of magnitude higher than the gravity forces. In this regime, the cohesiveness of the powder is expected to have considerable influence on the characteristics (e.g. packing fraction) of the powder layer~\citep{Yang2000}.\\

\begin{figure}[h!!]
   \centering
   \subfigure[Simulation model for the powder recoating process in metal AM: Initial configuration, $\gamma=\gamma_0$.]
   {
    \includegraphics[width=0.8\textwidth]{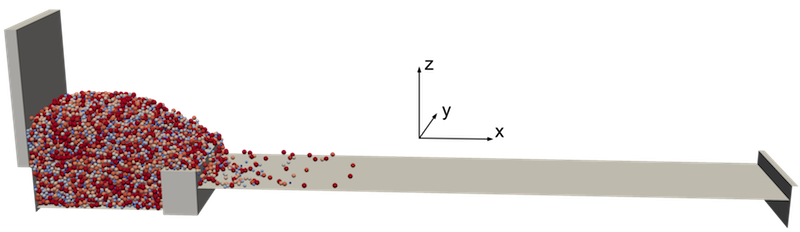}
    \label{fig:model_configs1}
   }
   \centering
   \subfigure[Intermediate configuration of powder recoating process simulation for the case $\gamma=\gamma_0$.]
   {
    \includegraphics[width=0.8\textwidth]{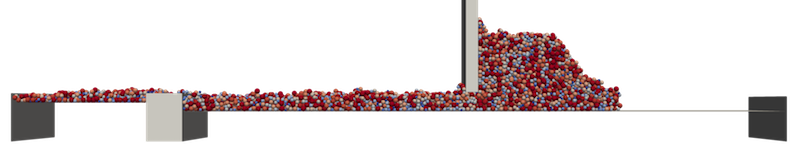}
    \label{fig:model_configs2}
   }
   \centering
   \subfigure[Intermediate configuration of powder recoating process simulation for the case $\gamma=0$.]
   {
    \includegraphics[width=0.8\textwidth]{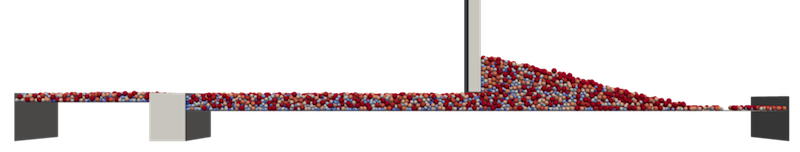}
    \label{fig:model_configs3}
   }
   \caption{Different configurations of the employed simulation model for the powder recoating process in metal AM.}
  \label{fig:model_configs}
\end{figure}

The geometry of the powder recoating model in its initial configuration is shown in Figure~\ref{fig:model_configs1}. First, a global Cartesian coordinate frame is defined, with the x-axis pointing in recoating direction, and the z-axis representing the build platform normal direction (see Figure~\ref{fig:model_configs1}). In order to generate the illustrated powder particle configuration, powder particles with random size distribution are initially placed on a Cartesian grid. Under the action of gravity, the particles fall down and finally settle in a static equilibrium configuration similar to Figure~\ref{fig:model_configs1}. Especially for cohesive powders, a realistic gravity-driven initial arrangement of the particles is important, since this first step already models a possible pre-compaction (due to gravity and inertia forces), which determines the subsequent bulk powder behavior. After the particles have settled, a defined amount of powder is provided for the recoating process by driving the platform of the powder reservoir upwards. Subsequently, the recoating blade is driven from the left to the right with a defined, constant velocity $V_B$ to finally spread the powder across the bottom of the powder bed, also denoted as substrate in the following. The standard blade velocity is chosen as $V_B=V_0=10mm/s$, unless stated otherwise. In this low velocity regime, inertia effects can be neglected and the spreading process can be considered as quasi-static. On the one hand, this strategy helps to isolate the effects under consideration (e.g. adhesion, layer thickness etc.) from the potential influence of inertia effects. On the other hand, it provides an upper bound for the achievable powder layer quality, which typically decreases with increasing spreading dynamics~\citep{Mindt2016,Parteli2016}. In Section~\ref{sec:recoating_bladevelocity}, the influence of blade velocity will be investigated separately at otherwise fixed process parameters.\\

In the present setup, the nominal powder layer thickness $t_0$ is defined as the vertical distance from the bottom of the powder bed to the upper edges of the left / right powder bed boundaries. There is a defined gap of $d_{min}/2=10\mu m$ between the lower edge of the recoating blade and these two upper edges of the left / right powder bed boundaries, which leads to a better agreement of nominal layer height $t_0$ and real mean layer height $t$. The powder bed dimensions in x- and y-direction are fixed to $5x1mm$, and periodic boundary conditions are applied in $y-$direction. While different layer thicknesses $t_0=d_{max,0}, \,\, 2d_{max,0},\,\, 3d_{max,0},\,\, 4d_{max,0}$ will be investigated in Section~\ref{sec:recoating_thickness}, all remaining studies are based on a fixed layer thickness of $t_0=3d_{max,0}$. In order to still guarantee for a small gap between recoating blade and the largest powder particles in the worst case $t_0=d_{max,0}$, the "nominal" maximal particle diameter $d_{max,0}=50 \mu m$ has been chosen slightly larger than the "actual" maximal particle diameter $d_{max}=D90=44 \mu m$ (see Section~\ref{sec:model}). The present work will exclusively focus on stiff / rigid recoating blades of rectangular shape as illustrated in Figure~\ref{fig:model_configs}. The analysis of alternative recoating tools such as soft / compliant blades, rakes or rollers is subject to future studies.\\

Figures~\ref{fig:model_configs2} and~\ref{fig:model_configs3} show intermediate configurations of the recoating process for the adhesion values $\gamma=\gamma_0$ and $\gamma=0$, respectively. Accordingly, the consideration of adhesion leads to a more realistic bulk powder behavior without particles rolling across the bottom of the powder bed. Further, it leads to a less uniform / smooth powder layer surface, to rather irregular and steeper slopes of the powder pile, to powder particles that stick to the recoating blade, and also to avalanching effects of cohesive powder agglomerates during the spreading process. These observations are well-known from experimental studies of the powder recoating process~\citep{Ebert2003,Tan2017,Yablokova2015}, and shall in the following be studied and quantified numerically. Therefore, two proper powder layer metrics will be defined in the subsequent section.\\

\subsection{Metrics for powder bed characterization}
\label{sec:recoating_metrics}

\begin{figure}[h!!]
    \centering
   \subfigure[Definition of surface profile.]
   {
    \includegraphics[width=0.3\textwidth]{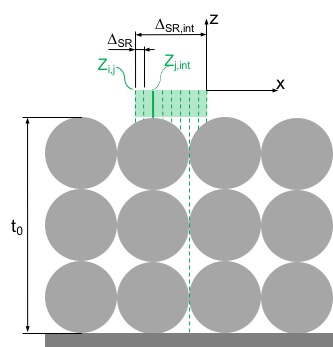}
    \label{fig:metrics_sr}
   }
   \hspace{2.0 cm}
 \centering
 \subfigure[Definition of packing fraction.]
   {
    \includegraphics[width=0.3\textwidth]{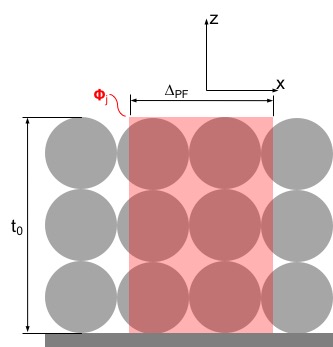}
    \label{fig:metrics_pf}
   }
   \caption{Schematic diagrams for the definition of surface profile and packing fraction of powder layer.}
  \label{fig:metrics}
\end{figure}

Before studying different powder layer configurations, proper metrics need to be defined to quantify and assess the layer characteristics. Throughout this work, two metrics, namely the surface profile and the packing fraction of the powder layer, will be employed for this purpose (see Figure~\ref{fig:metrics}). In order to measure the surface profile, vertical lines / rays are defined on a Cartesian grid with resolution $\Delta_{SR}$ in the xy-plane. For each ray, all intersection points with neighboring particles are determined, and the maximal z-coordinate of these intersection points is taken as surface profile height $z$ at this xy-position. If no intersection point with neighboring particles exists, the z-coordinate of the bottom of the powder bed is taken. In a second step, a coarser Cartesian grid with resolution $\Delta_{SR,int}>\Delta_{SR}$ is introduced, and the integrated / filtered surface profile height $z_{j,int}$ is defined for each 2D segment $j$ on this coarser grid as the maximal profile height of all rays $i$ within this segment, i.e. $z_{j,int}:=\max_{i}(z_{i,j})$ (see Figure~\ref{fig:metrics_sr}). In this context, the mean surface profile height is defined as $t:=<\!z_{j,int}\!>$. Here and in the following, the operators $<...>$ and $std(...)$ represent the mean and the standard deviation of a (discrete) spatial field. As second metric, the particle packing fraction is considered. In order to also analyze spatial variations of the packing fraction in the sense of a 2D xy-field, a further Cartesian grid with resolution $\Delta_{PF}$ is introduced, and the local packing fraction value within each 3D bin spanned by the 2D segment $j$ on this grid and the nominal powder layer thickness $t_0$ is defined as the ratio of particle volume to bin volume, i.e. $\Phi_{t0,j}:=V_{p,j}/(t_0 \Delta_{PF}^2)$. Equivalently, the packing fraction can also be defined within a powder layer volume confined by the (actual) mean powder layer thickness $t$, i.e. $\Phi_{t,j}:=V_{p,j}/(t \Delta_{PF}^2)$. Certain process parameter choices (see e.g. Section~\ref{sec:recoating_bladevelocity}) may lead to a considerably decreased mean layer height $t$ going along with an (almost) unchanged packing fraction level. Under such circumstances, the (frequently employed) definition $\Phi_{t0}$ based on the nominal layer height $t_0$ would lead to a considerably distorted result. Therefore, the definition $\Phi_{t}$ will be preferred throughout this work. The accumulated particle volume within one bin is calculated via numerical integration (in order to account for the volume of particles cut by the bin walls with sufficient accuracy) based on a simple voxel strategy with cubical volume elements of side length $\Delta_{V}$. This volume discretization length has to be small compared to the smallest particle diameter $d_{min}$ to ensure sufficient integration accuracy. In the present study, this (purely numerical) parameter has been taken as $\Delta_{V}=d_{min}/8=2.5 \mu m$ resulting in relative packing fraction errors in the range of $0.1\%$.\\

In the following, the definition of these two metrics shall briefly be motivated. The grid with resolution $\Delta_{PF}$ has been introduced to enable the calculation of a spatial packing fraction field $\Phi(x,y)$, which in turn allows to quantify the variation of the packing fraction across the powder bed. It is important to note that the choice of the resolution / bin size $\Delta_{PF}$ crucially influences the resulting packing fraction variation. From Figure~\ref{fig:metrics_pf}, it becomes obvious that a maximal variation between the values $\Phi=1$ and $\Phi=0$ might result for very small bin sizes $\Delta_{PF}/d_{min} \ll 1$ even if a perfectly uniform packing is considered. Thus, the relation $\Delta_{PF}/d_{min} > 1$ can be regarded as minimal required for the bin size. On the other hand, when the bin size approaches the dimensions of the powder layer, the variation decreases to zero independent of the actual powder bed characteristic. Consequently, this bin size should be motivated physically, i.e. in dependence on the question that has to be answered by the metric packing fraction variation. In the present study, the bin size has been set to $\Delta_{PF}=100 \mu m$, which is in the range of a typical laser beam diameter employed in SLM processes. This choice can be justified as follows: On the one hand, packing fraction variations on smaller length scales are only of secondary interest, since these might be homogenized by the highly dynamic material flow within the melt pool. On the other hand, variations on the length scale of the laser beam diameter and above might indeed lead to an off-set of adjacent melt tracks and should be resolved by a reasonable metric.\\

\begin{figure}[h!!]
    \centering
   \subfigure[Low SR, High PF.]
   {
    \includegraphics[width=0.22\textwidth]{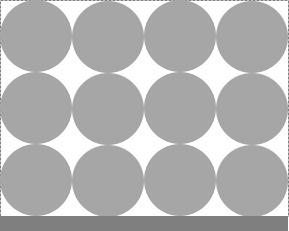}
    \label{fig:metrics2_1}
   }
   \hspace{0.2 cm}
 \centering
 \subfigure[Low SR, Low PF.]
   {
    \includegraphics[width=0.22\textwidth]{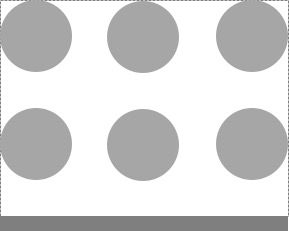}
    \label{fig:metrics2_2}
   }   
   \hspace{0.2 cm}
   \centering
   \subfigure[High SR, High PF.]
   {
    \includegraphics[width=0.22\textwidth]{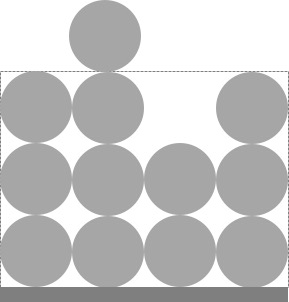}
    \label{fig:metrics2_3}
   }
   \hspace{0.2 cm}
 \centering
 \subfigure[High SR, Low PF.]
   {
    \includegraphics[width=0.22\textwidth]{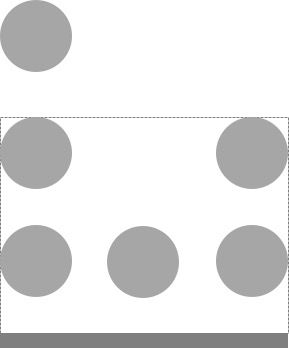}
    \label{fig:metrics2_4}
   }
   \caption{Examples for powder particle configurations leading to low / high surface roughness (SR) as well as high / low packing fraction (PF).}
  \label{fig:metrics2}
\end{figure}

Throughout this work, the variation of the surface profile height will be considered as metric for surface uniformity and denoted as surface roughness of the powder layer (in contrast to the surface roughness of individual powder particles due to surface asperities on the nanoscale). The integrated approach of measuring surface roughness as described above can mainly be motivated by three arguments. First, measuring the surface profile only on the basis of the rays $i$ without the subsequent integration / filter step accross the segments $j$ is equivalent to a mechanical measurement of the surface profile with a vertical rod of vanishing diameter. On the other hand, the proposed approach is equivalent to mechanically measuring the surface profile with a rod of finite thickness $\Delta_{SR,int}$, leading to a limited / finite resolution of the surface roughness, which is equivalent to common (non-mechanical) approaches of experimental surface profile characterization~\citep{Neef2014}. Second, a minimal / optimal surface roughness is achieved if the particles are arranged on a regular grid in a manner as shown in Figure~\ref{fig:metrics2_1}. The proposed metric naturally assigns a minimal surface roughness value of zero to this optimal configuration. Third, and most important, the proposed metric leads to a pure surface characterization and isolates the effects of surface roughness and packing fraction, with the latter representing a metric of volumetric powder characterization. This can be explained by considering the four different particle configurations illustrated in Figure~\ref{fig:metrics2}. In realistic powder beds, there is always a (comparatively high) number of rays reaching the substrate without intersecting any particle. This effect would lead to surface roughness values, if only defined via "unfiltered" rays, that is strongly dependent on the powder layer thickness. The thickness of the powder layer should, however, not influence the value of a properly defined metric for surface characerization. Moreover, this effect would e.g. also lead to considerably different surface roughness values between the configurations in Figures~\ref{fig:metrics2_1} and~\ref{fig:metrics2_2}, which actually rather differ in packing fraction than in surface uniformity. Specifically, the proposed definitions of (filtered / integrated) surface roughness and packing fraction allow to distinguish and isolate distinct configurations as illustrated in Figures~\ref{fig:metrics2_1} to~\ref{fig:metrics2_4}. In practice, this can be very useful since packing fraction (variations) can e.g. be associated with thermal conductivity (variations) in the powder bed~\citep{Gusarov2003,Lechman2013}, the effective / integrated absorptivity across the powder layer thickness with respect to incident laser energy~\citep{Boley2015,Gusarov2008}, or shrinkage and resulting layer height (variations) of the solidified material. On the other hand, surface roughness in the sense of layer height non-uniformity on the length scale of individual particles might lead to an inhomogeneous energy supply resulting in local overheating / evaporation or partly molten particles / inclusions, and to an increased tendency of particles on the top of local surface elevations to be dragged away by the gas flow in the build chamber~\citep{Matthews2016,Zhirnov2018}.\\

\begin{figure}[h!!]
    \centering
   \subfigure[Packing fraction $\Phi(x,y)$ of the case  $\gamma=\gamma_0$.]
   {
    \includegraphics[width=0.48\textwidth]{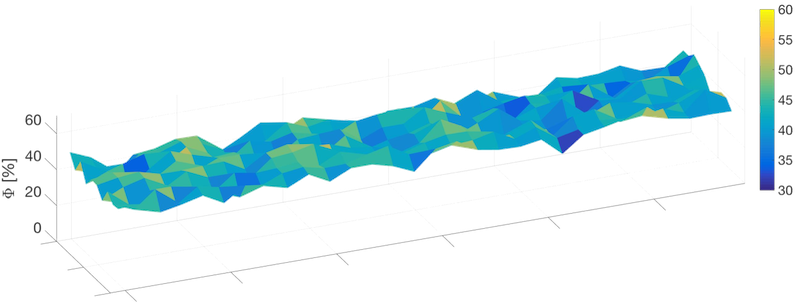}
    \label{fig:metrics3_1}
   }
    \hspace{0.1 cm}
   \centering
   \subfigure[Packing fraction $\Phi(x,y)$ of the case  $\gamma=0$.]
   {
    \includegraphics[width=0.48\textwidth]{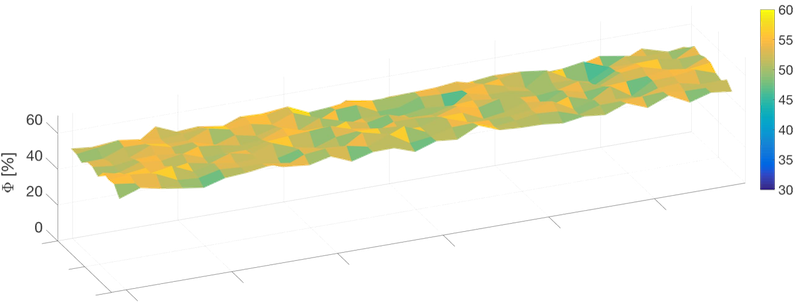}
    \label{fig:metrics3_2}
   }
   \hspace{0.1 cm}
 \centering
 \subfigure[Surface profile $z_{int}(x,y)$ of the case  $\gamma=\gamma_0$.]
   {
    \includegraphics[width=0.48\textwidth]{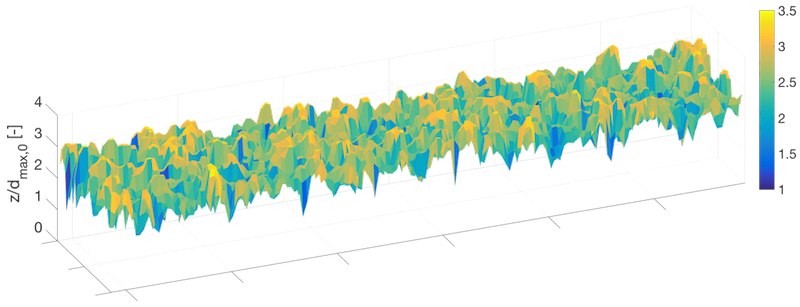}
    \label{fig:metrics3_3}
   }   
   \hspace{0.1 cm}
 \centering
 \subfigure[Surface profile $z_{int}(x,y)$ of the case  $\gamma=0$.]
   {
    \includegraphics[width=0.48\textwidth]{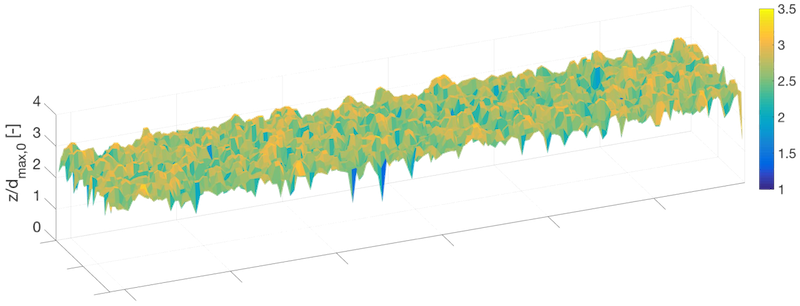}
    \label{fig:metrics3_4}
   }
   \caption{Spatial packing fraction  $\Phi(x,y)$ and surface profile $z_{int}(x,y)$ fields for the cases $\gamma=\gamma_0$ and $\gamma=0$ ($t_0=3d_{max,0}$).}
  \label{fig:metrics3}
\end{figure}

In order to avoid the undesirable influence of boundary effects, the statistical powder bed evaluation will only be conducted within an area of $3x1mm$ at the center of the powder bed. In Figure~\ref{fig:metrics3}, the spatial fields of packing fraction $\Phi(x,y)$ (top row) and surface profile $z_{int}(x,y)$ (bottom row) are plotted for a cohesive powder ($\gamma=\gamma_0$, left column) and a non-cohesive powder ($\gamma=0$, right column). Accordingly, the cohesive powder results in a decreased powder bed quality characterized by an increased spatial variation of packing fraction and surface profile height. Moreover, it appears that the corresponding mean values are lower for the cohesive powder. In the following sections, these observations will be quantified numerically for different choices of the process parameters. Finally, it shall be noted that throughout this work, only "first-layer" simulations are conducted, which helps to reduce the complexity in interpreting results and to isolate the parameters under consideration from potential effects due to interaction with previous powder layers. However, the important case of powder being spread on (molten and) solidified material of the previous layer will be mimicked by analyzing varied interaction parameters (e.g. surface energy, friction coefficient) between powder and substrate (see Section~\ref{sec:recoating_bladeandsubstrate}).

\subsection{Sensitivity with respect to stochastic variation in powder size distribution}
\label{sec:recoating_variation}

\begin{figure}[h!!]
   \centering
   \subfigure[Mean value of packing fraction field $\Phi_t(x,y)$.]
    {
    \includegraphics[width=0.48\textwidth]{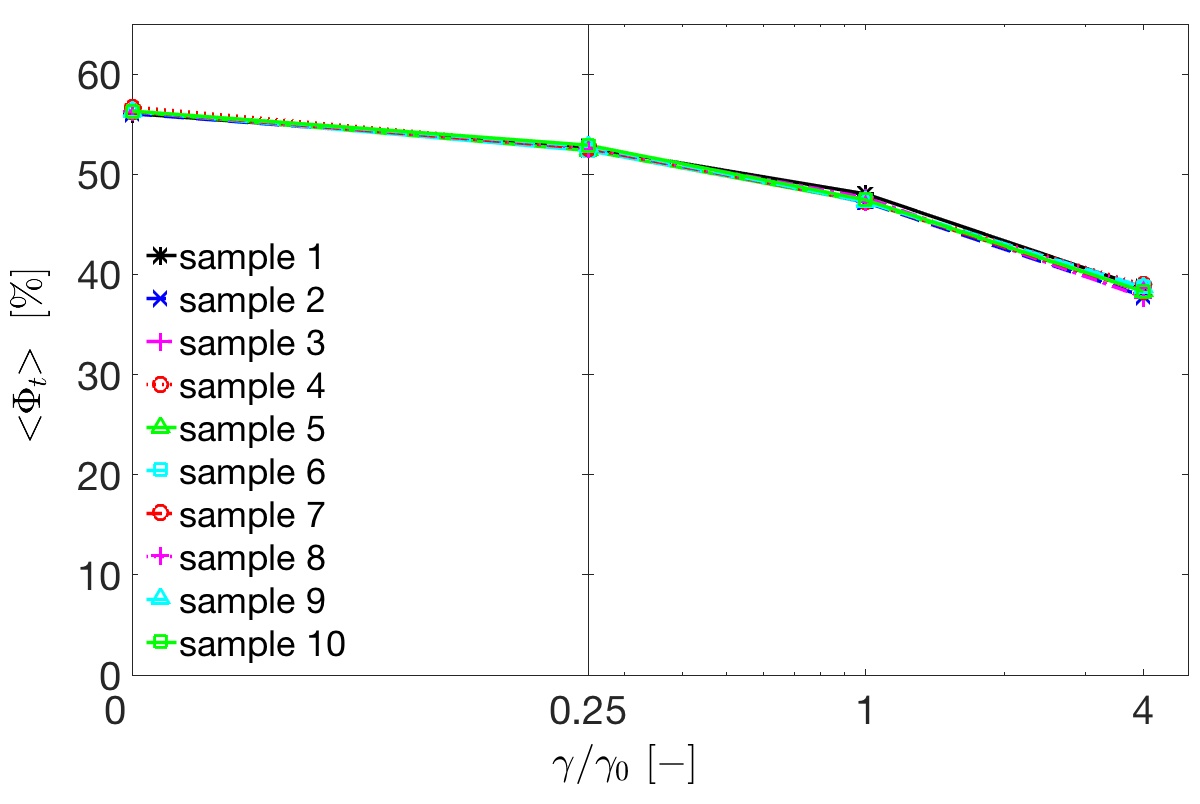}
    \label{fig:variation_1}
   }
  \hspace{0.1 cm}
   \centering
 \subfigure[Mean value of surface profile field $z_{int}(x,y)$.]
   {
    \includegraphics[width=0.48\textwidth]{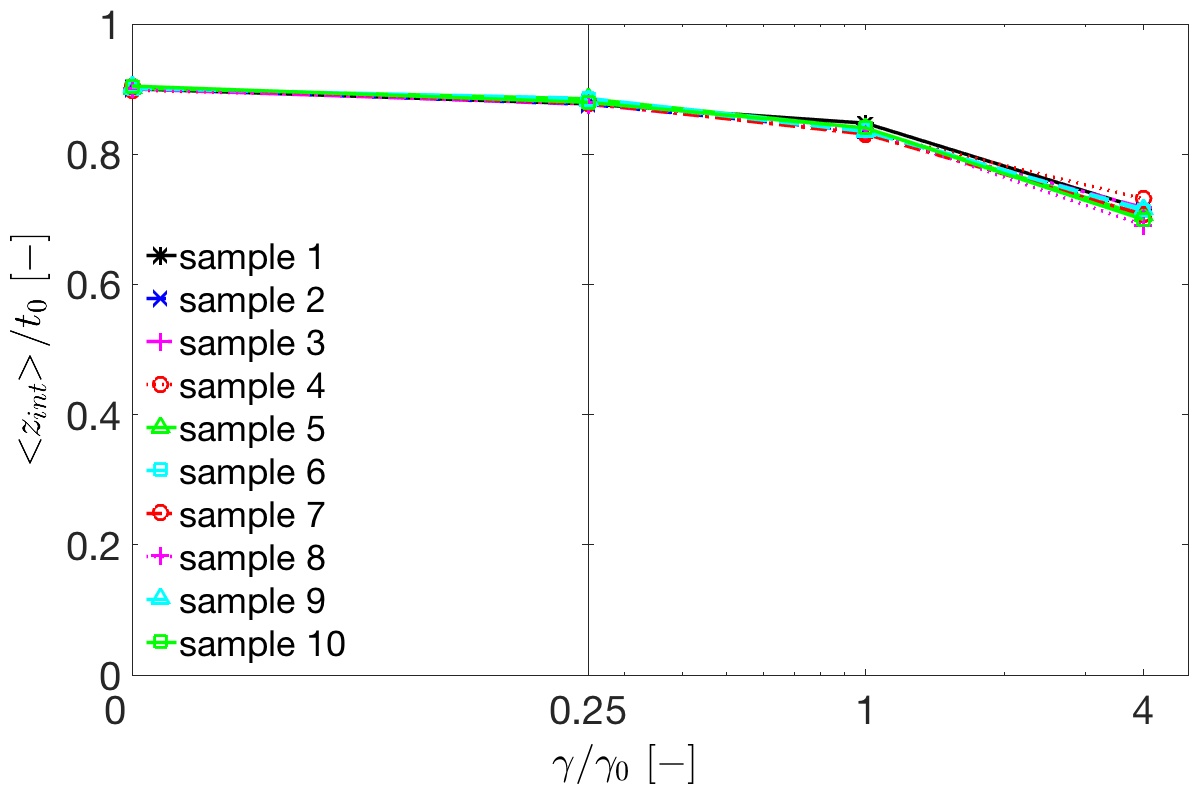}
    \label{fig:variation_2}
   }    
   \centering
    \subfigure[Standard deviation of packing fraction field $\Phi_t(x,y)$.]
    {
    \includegraphics[width=0.48\textwidth]{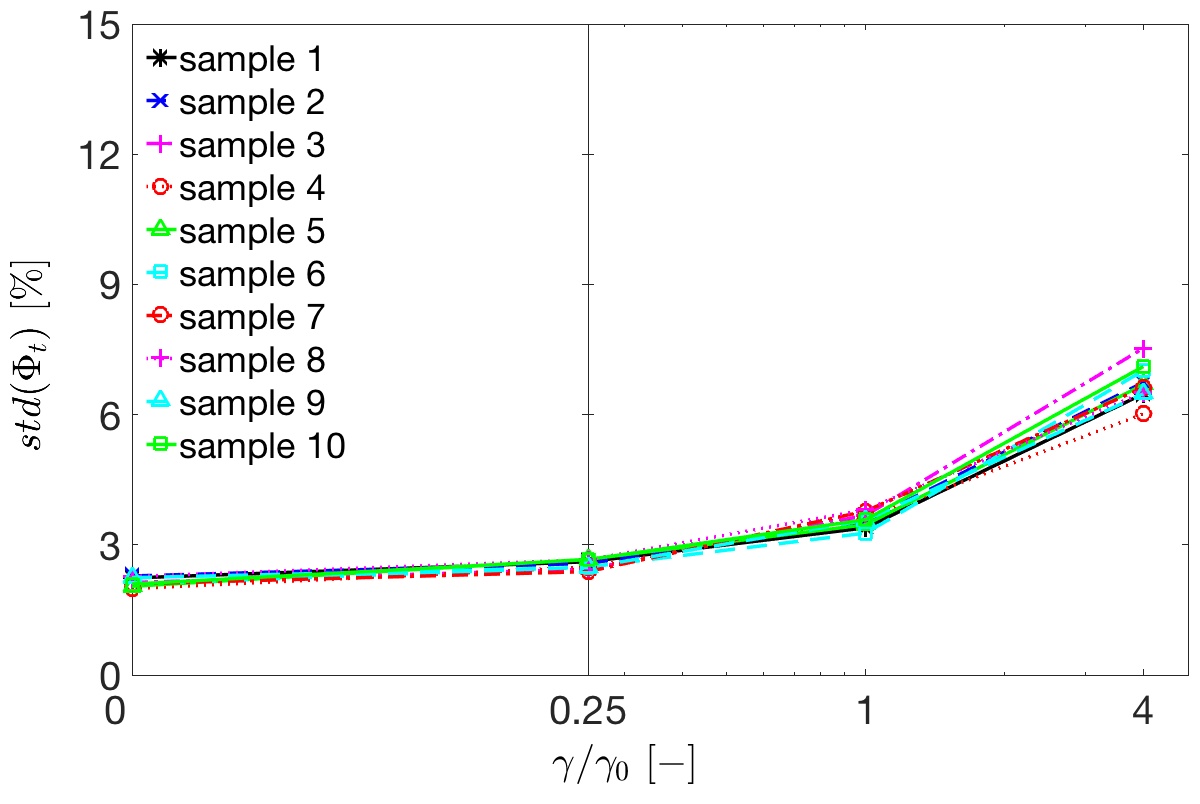}
    \label{fig:variation_3}
   }
   \hspace{0.1 cm}
 \centering
 \subfigure[Standard deviation of surface profile field $z_{int}(x,y)$.]
   {
    \includegraphics[width=0.48\textwidth]{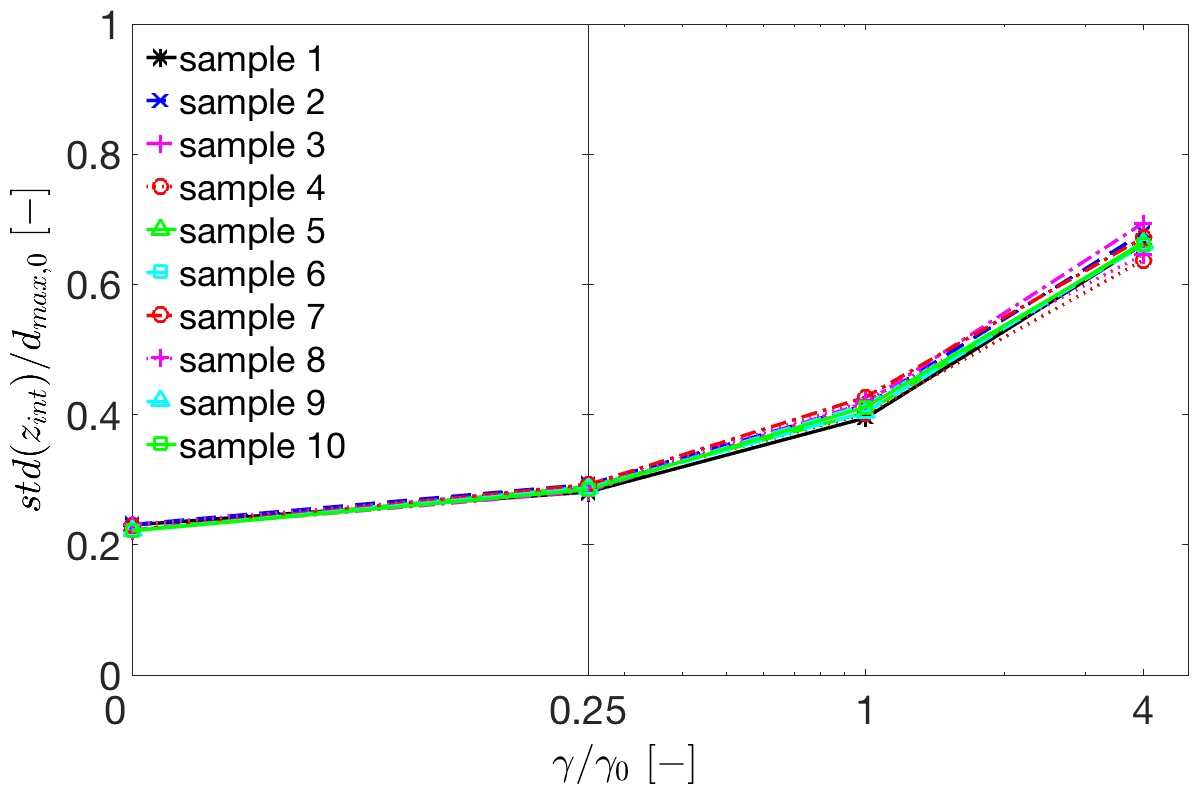}
    \label{fig:variation_4}
   }
   \caption{Mean value and standard deviation of $\Phi_t(x,y)$ and $z_{int}(x,y)$ as function of powder cohesiveness $\gamma / \gamma_0$ for ten realizations of the powder particle size distribution ($t_0=3d_{max,0}$).}
  \label{fig:variation}
\end{figure}

As already stated above, the powder particle size distribution is modeled as a log-normal distribution. In this section, the sensitivity of the resulting powder layer characteristics with respect to different numerical / stochastic realizations of this (fixed) log-normal distribution (i.e. when taking different powder samples of the same size class) will be investigated. Figure~\ref{fig:variation} illustrates the mean values and standard deviations of the packing fraction $\Phi_t(x,y)$ and surface profile height $z_{int}(x,y)$ field for ten powder samples / numerical realizations of the considered powder particle size distribution. All in all, it can be concluded that the deviations between the individual curves are small and, consequently, the sensitivity with respect to different powder samples is reasonably small as well. Assuming that the averaging across different powder samples is equivalent to the averaging across an increased spatial domain, this observation also confirms that the size of the representative powder layer volume has been chosen sufficiently large in this study. Only in the range of the highest adhesion value $\gamma=4\gamma_0$, the deviations between the different powder samples are slightly increased, especially when considering the spatial variations of $\Phi_t(x,y)$ and $z_{int}(x,y)$ (bottom row). This observation might be explained by the presence of larger cohesive particle agglomerates, which would require a larger representative volume to achieve similarly small deviations as for the non-cohesive powder. Since the variance across the different powder samples is still reasonably small compared to the change in the considered metrics when comparing different surface energy values, this effect will not be further analyzed in the following. However, if required, this variance could easily be reduced by either considering a larger representative volume or by averaging results across different powder samples.\\

Apart from the effect of different powder samples, the general trends exemplarily observed in Figure~\ref{fig:metrics3} can be confirmed: Increasing surface energy / cohesiveness leads to decreasing powder layer quality. Concretely, the mean packing fraction $<\!\Phi_t(x,y)\!>$ decreases from almost $60\%$ to a value below $40\%$, and the mean layer height $t:=<\!z_{int}(x,y)\!>$ decreases from $90\%$ to approximately $70\%$ of the nominal layer height $t_0$ when increasing the surface energy from $\gamma=0$ to $\gamma=4\gamma_0$. Similarly, the standard deviation of the packing fraction $std(z_{int}(x,y))$ increases from $2.5\%$ to $7.5\%$, and the standard deviation of the surface profile height $std(z_{int}(x,y))$ increases from approximately $20\%$ to a value above $60\%$ of the (nominal) maximal particle diameter $d_{max,0}$. In Section~\ref{sec:recoating_thickness}, these observations will be discussed in further detail.\\

\subsection{Sensitivity with respect to the choice of mechanical powder properties}
\label{sec:recoating_powderproperties}

While the employed surface energy value $\gamma_0$ has been determined for the specific powder material based on a fitting of numerical and experimental AOR measurements, standard values from the literature have been taken for other powder material parameters such as the friction coefficient $\mu$ or the coefficient of restitution $c_{COR}$. Moreover, under the common assumption that global powder kinematics remain unaffected, the stiffness / penalty parameter $k_N$ has been chosen by one to two orders of magnitude smaller than the Young's modulus of Ti-6Al-4V~\citep{Meier2018a}. The following sensitivity analysis shall justify the choice of these parameters by comparing the powder layer characteristics resulting from this standard parameter set and modified parameter values. Specifically, the friction coefficient and the coefficient of restitution are increased by $50\%$, the stiffness parameter is increased by a factor of four. In all cases, identical parameters are chosen for particle-to-particle and particle-to-wall interaction.\\

\begin{figure}[h!!]
   \centering
   \subfigure[Mean value of packing fraction field $\Phi_t(x,y)$.]
    {
    \includegraphics[width=0.48\textwidth]{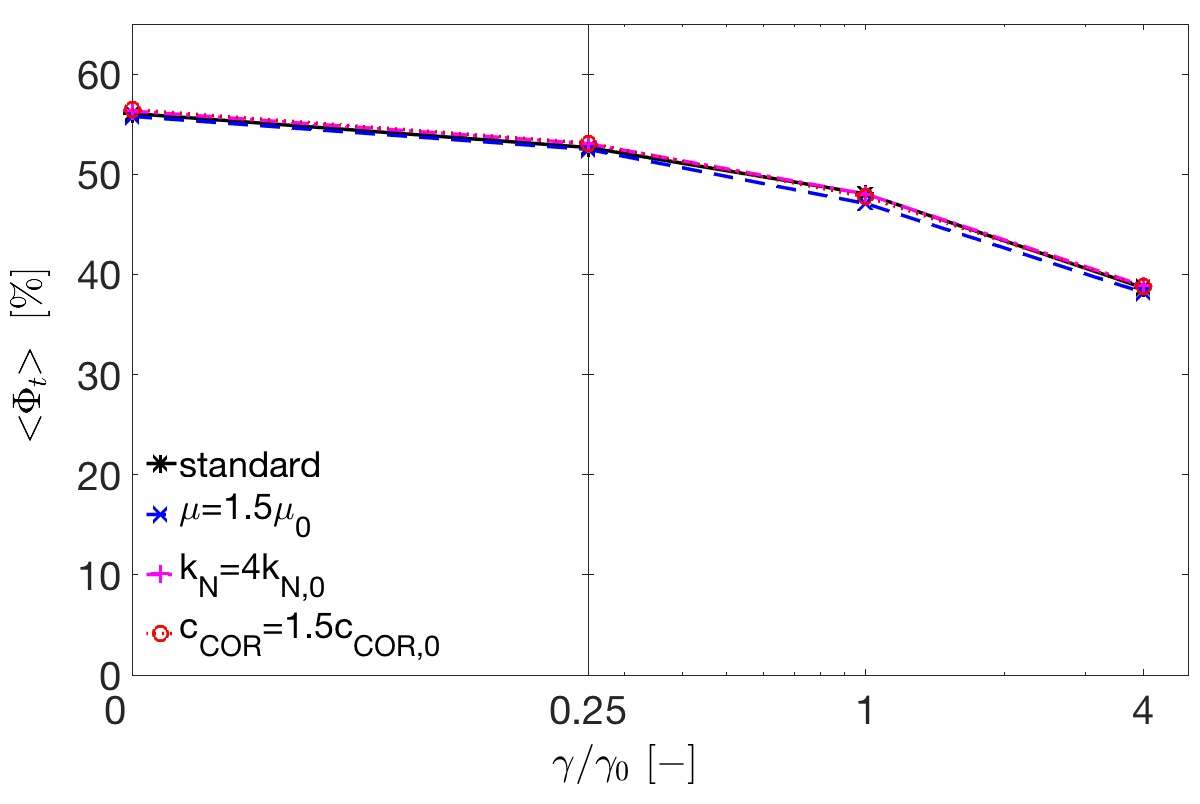}
    \label{fig:parameter_1}
   }
  \hspace{0.1 cm}
   \centering
 \subfigure[Mean value of surface profile field $z_{int}(x,y)$.]
   {
    \includegraphics[width=0.48\textwidth]{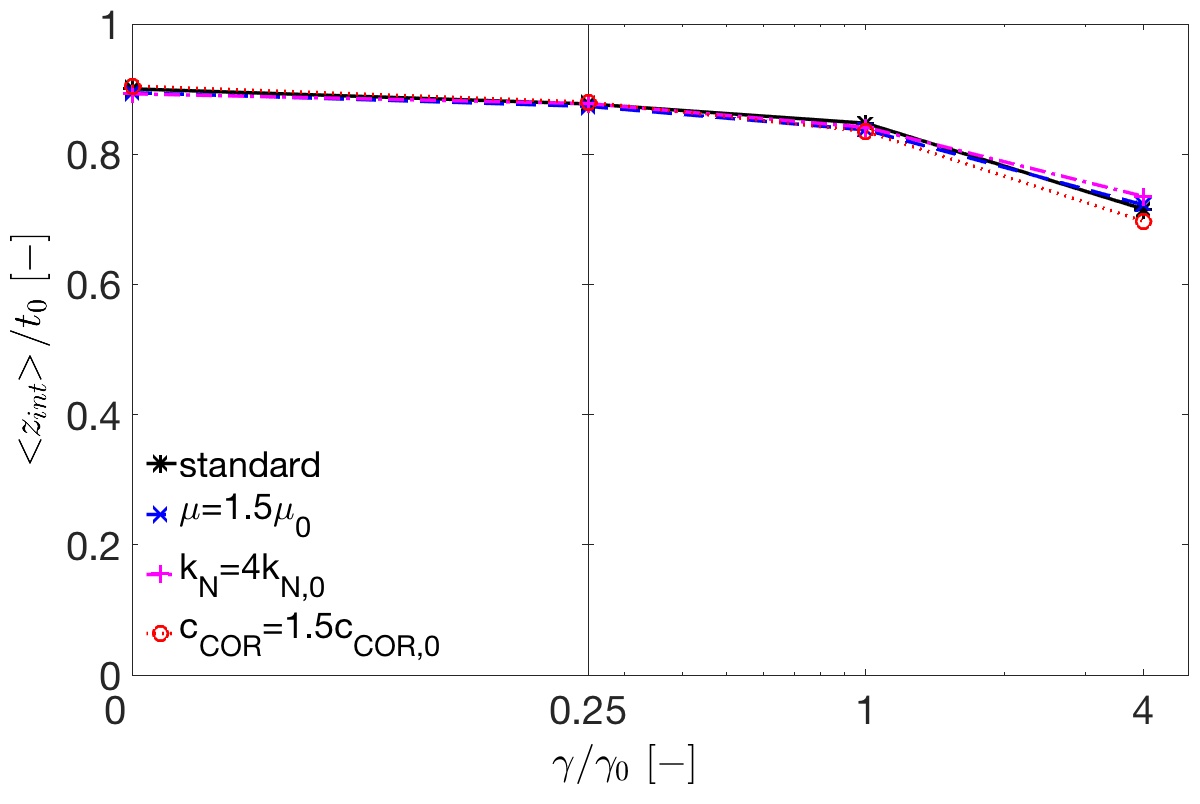}
    \label{fig:parameter_2}
   }    
   \centering
    \subfigure[Standard deviation of packing fraction field $\Phi_t(x,y)$.]
    {
    \includegraphics[width=0.48\textwidth]{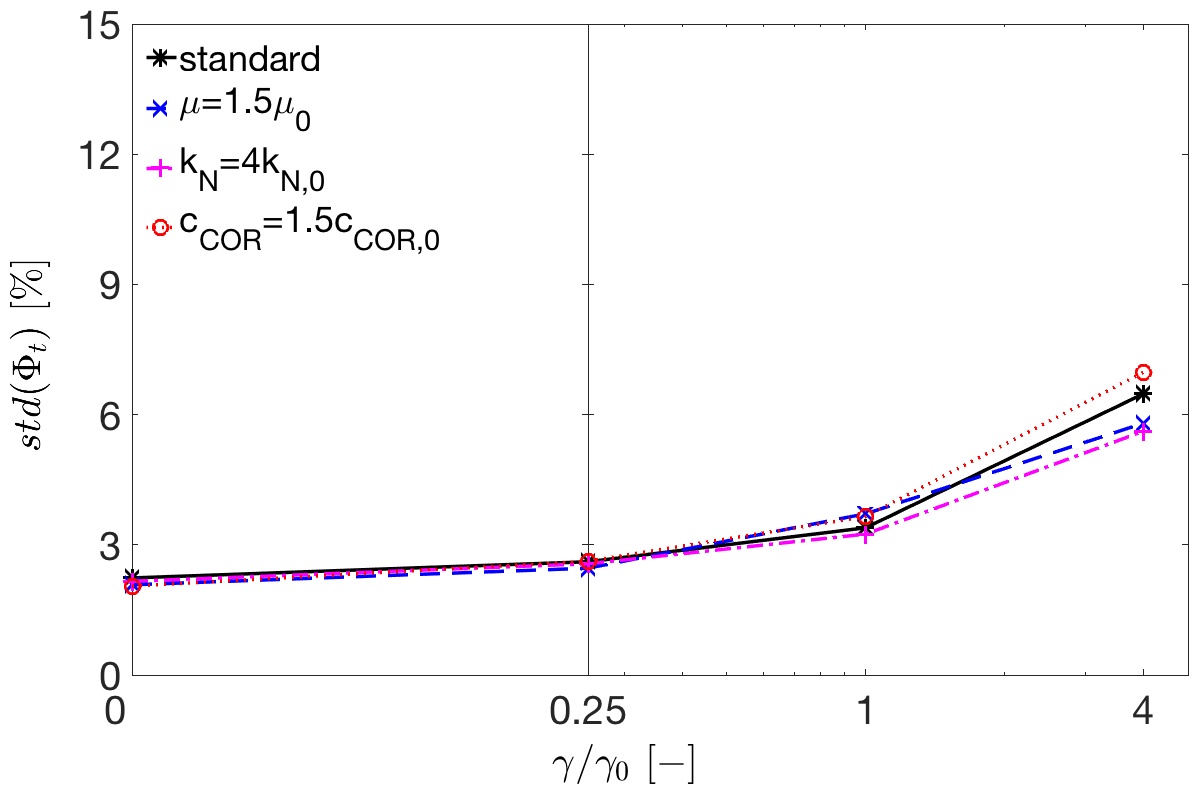}
    \label{fig:parameter_3}
   }
   \hspace{0.1 cm}
 \centering
 \subfigure[Standard deviation of surface profile field $z_{int}(x,y)$.]
   {
    \includegraphics[width=0.48\textwidth]{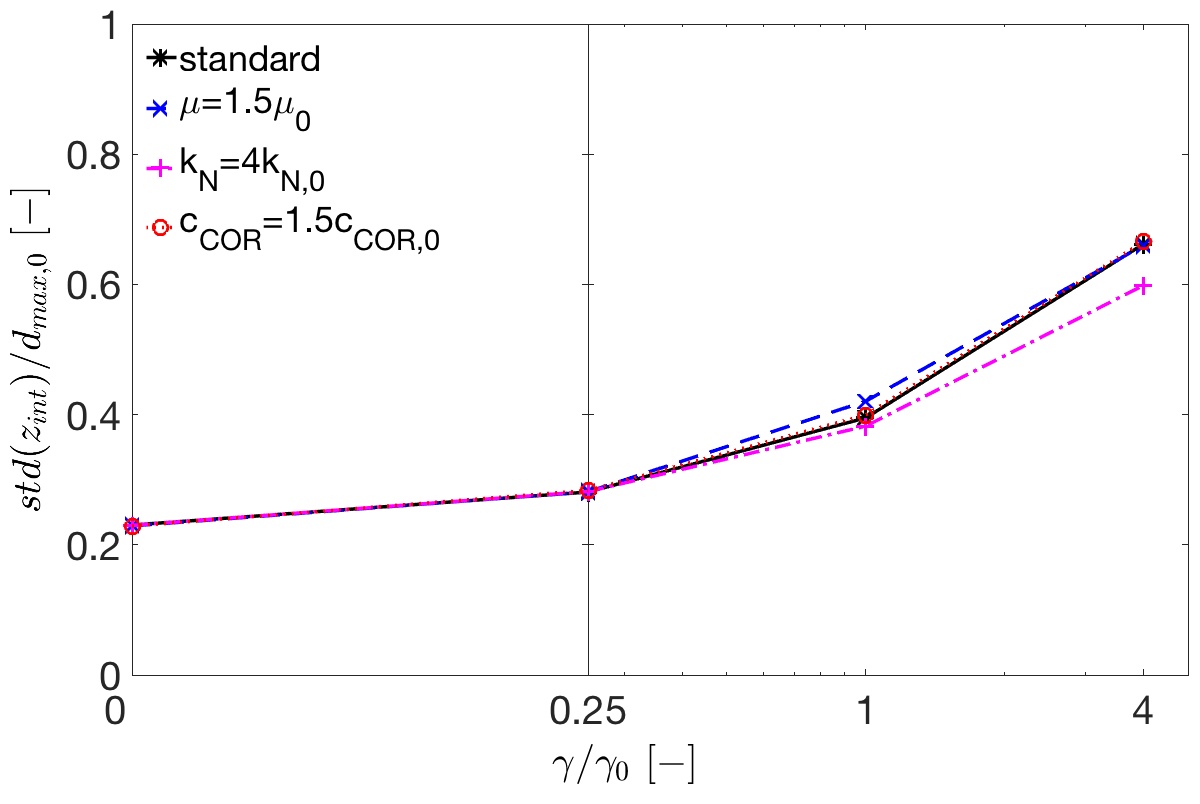}
    \label{fig:parameter_4}
   }
   \caption{Mean value and standard deviation of $\Phi_t(x,y)$ and $z_{int}(x,y)$ as function of powder cohesiveness $\gamma / \gamma_0$ for different choices of the parameters $\mu, \, k_N$ and $c_{COR}$ ($t_0=3d_{max,0}$).}
  \label{fig:parameter}
\end{figure}

According to Figure~\ref{fig:parameter}, the sensitivity / uncertainty of the considered powder layer metrics with respect to these parameters is again small compared to the increments between the different surface energy values. Moreover, also the slightly increased deviations observed for the highest surface energy value $\gamma=4\gamma_0$ are in the range of the uncertainty due to stochastic powder size variations as discussed in the last section. Consequently, the uncertainty in the choice of these parameters seems to have no noticeable effect on the general statements made in the following sections.

\subsection{Influence of powder layer thickness}
\label{sec:recoating_thickness}

\begin{figure}[h!!]
    \centering
   \subfigure[$t_0\!=\!d_{max,0}$, \,\,\,\, $\gamma\!=\!0$.]
   {
    \includegraphics[width=0.46\textwidth]{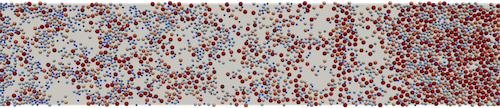}
    \label{fig:tv_1}
   }
    \hspace{0.1 cm}
    \subfigure[$t_0\!=\!d_{max,0}$, \,\,\,\, $\gamma\!=\!\gamma_0/4$.]
    {
    \includegraphics[width=0.46\textwidth]{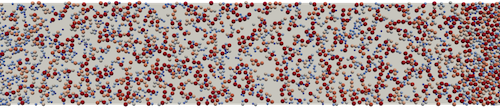}
    \label{fig:tv_2}
   }
    \subfigure[$t_0\!=\!d_{max,0}$, \,\,\,\, $\gamma\!=\!\gamma_0$.]
       {
    \includegraphics[width=0.46\textwidth]{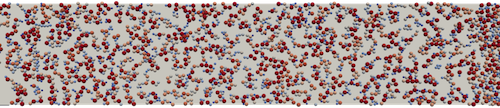}
    \label{fig:tv_3}
   }
    \hspace{0.1 cm}
    \subfigure[$t_0\!=\!d_{max,0}$, \,\,\,\, $\gamma\!=\!4\gamma_0$.]
       {
    \includegraphics[width=0.46\textwidth]{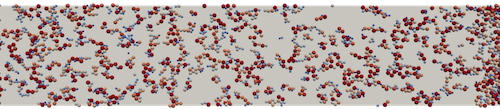}
    \label{fig:tv_4}
   }
    \subfigure[$t_0\!=\!2d_{max,0}$, \,\,\,\, $\gamma\!=\!0$.]
       {
    \includegraphics[width=0.46\textwidth]{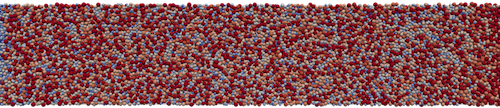}
    \label{fig:tv_5}
   }
    \hspace{0.1 cm}
    \subfigure[$t_0\!=\!2d_{max,0}$, \,\,\,\, $\gamma\!=\!\gamma_0/4$.]
       {
    \includegraphics[width=0.46\textwidth]{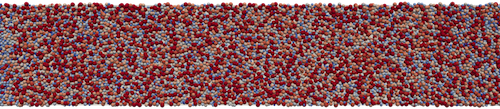}
    \label{fig:tv_6}
   }
    \subfigure[$t_0\!=\!2d_{max,0}$, \,\,\,\, $\gamma\!=\!\gamma_0$.]
       {
    \includegraphics[width=0.46\textwidth]{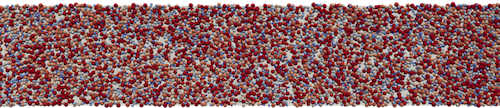}
    \label{fig:tv_7}
   }
    \hspace{0.1 cm}
    \subfigure[$t_0\!=\!2d_{max,0}$, \,\,\,\, $\gamma\!=\!4\gamma_0$.]
       {
    \includegraphics[width=0.46\textwidth]{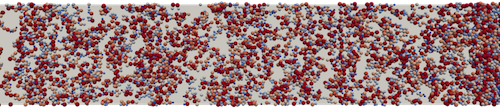}
    \label{fig:tv_8}
   }
   \subfigure[$t_0\!=\!3d_{max,0}$, \,\,\,\, $\gamma\!=\!0$.]
   {
    \includegraphics[width=0.46\textwidth]{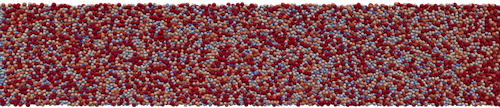}
    \label{fig:tv_9}
   }
    \hspace{0.1 cm}
    \subfigure[$t_0\!=\!3d_{max,0}$, \,\,\,\, $\gamma\!=\!\gamma_0/4$.]
    {
    \includegraphics[width=0.46\textwidth]{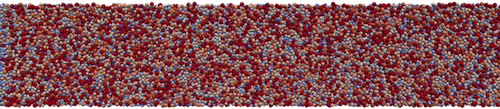}
    \label{fig:tv_10}
   }
    \subfigure[$t_0\!=\!3d_{max,0}$, \,\,\,\, $\gamma\!=\!\gamma_0$.]
       {
    \includegraphics[width=0.46\textwidth]{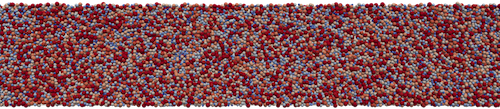}
    \label{fig:tv_11}
   }
    \hspace{0.1 cm}
    \subfigure[$t_0\!=\!3d_{max,0}$, \,\,\,\, $\gamma\!=\!4\gamma_0$.]
       {
    \includegraphics[width=0.46\textwidth]{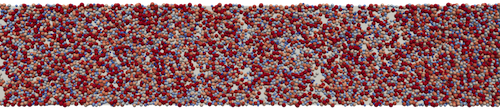}
    \label{fig:tv_12}
   }
    \subfigure[$t_0\!=\!4d_{max,0}$, \,\,\,\, $\gamma\!=\!0$.]
       {
    \includegraphics[width=0.46\textwidth]{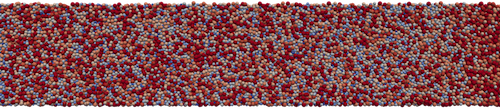}
    \label{fig:tv_13}
   }
    \hspace{0.1 cm}
    \subfigure[$t_0\!=\!4d_{max,0}$, \,\,\,\, $\gamma\!=\!\gamma_0/4$.]
       {
    \includegraphics[width=0.46\textwidth]{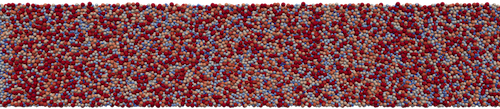}
    \label{fig:tv_14}
   }
    \subfigure[$t_0\!=\!4d_{max,0}$, \,\,\,\, $\gamma\!=\!\gamma_0$.]
       {
    \includegraphics[width=0.46\textwidth]{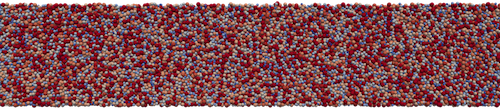}
    \label{fig:tv_15}
   }
    \hspace{0.1 cm}
    \subfigure[$t_0\!=\!4d_{max,0}$, \,\,\,\, $\gamma\!=\!4\gamma_0$.]
       {
    \includegraphics[width=0.46\textwidth]{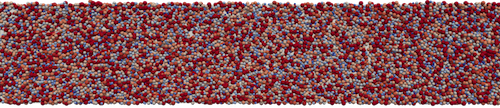}
    \label{fig:tv_16}
   }
   \caption{Resulting powder layers for different layer thicknesses $t_0$ and surface energies $\gamma$: Top view on powder bed.}
  \label{fig:tv}
\end{figure}

\begin{figure}[h!!]
    \centering
   \subfigure[$t_0\!=\!d_{max,0}$, \,\,\,\, $\gamma\!=\!0$.]
   {
    \includegraphics[width=0.46\textwidth]{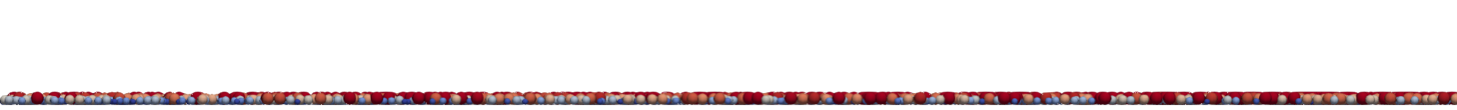}
    \label{fig:sv_1}
   }
    \hspace{0.1 cm}
    \subfigure[$t_0\!=\!d_{max,0}$, \,\,\,\, $\gamma\!=\!\gamma_0/4$.]
    {
    \includegraphics[width=0.46\textwidth]{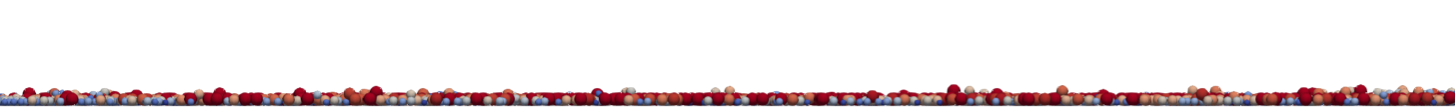}
    \label{fig:sv_2}
   }
    \subfigure[$t_0\!=\!d_{max,0}$, \,\,\,\, $\gamma\!=\!\gamma_0$.]
       {
    \includegraphics[width=0.46\textwidth]{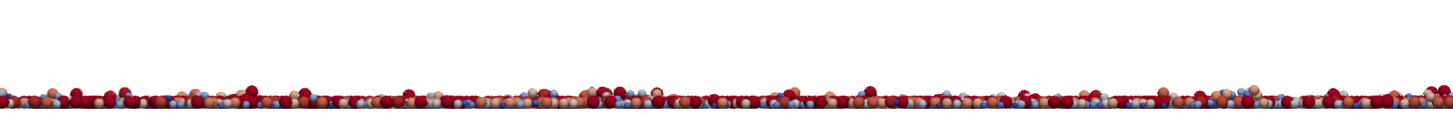}
    \label{fig:sv_3}
   }
    \hspace{0.1 cm}
    \subfigure[$t_0\!=\!d_{max,0}$, \,\,\,\, $\gamma\!=\!4\gamma_0$.]
       {
    \includegraphics[width=0.46\textwidth]{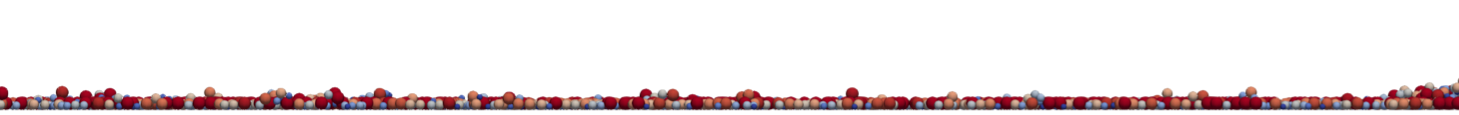}
    \label{fig:sv_4}
   }
    \subfigure[$t_0\!=\!2d_{max,0}$, \,\,\,\, $\gamma\!=\!0$.]
       {
    \includegraphics[width=0.46\textwidth]{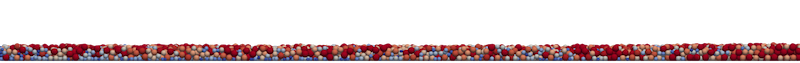}
    \label{fig:sv_5}
   }
    \hspace{0.1 cm}
    \subfigure[$t_0\!=\!2d_{max,0}$, \,\,\,\, $\gamma\!=\!\gamma_0/4$.]
       {
    \includegraphics[width=0.46\textwidth]{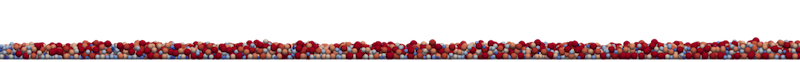}
    \label{fig:sv_6}
   }
    \subfigure[$t_0\!=\!2d_{max,0}$, \,\,\,\, $\gamma\!=\!\gamma_0$.]
       {
    \includegraphics[width=0.46\textwidth]{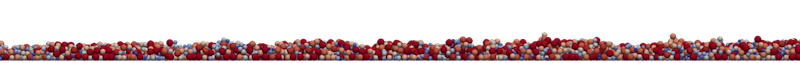}
    \label{fig:sv_7}
   }
    \hspace{0.1 cm}
    \subfigure[$t_0\!=\!2d_{max,0}$, \,\,\,\, $\gamma\!=\!4\gamma_0$.]
       {
    \includegraphics[width=0.46\textwidth]{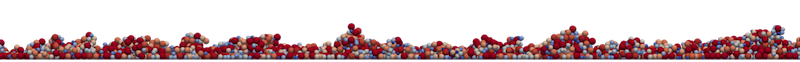}
    \label{fig:sv_8}
   }
   \subfigure[$t_0\!=\!3d_{max,0}$, \,\,\,\, $\gamma\!=\!0$.]
   {
    \includegraphics[width=0.46\textwidth]{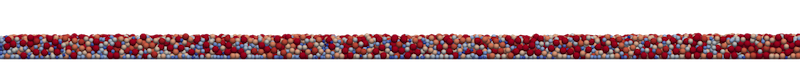}
    \label{fig:sv_9}
   }
    \hspace{0.1 cm}
    \subfigure[$t_0\!=\!3d_{max,0}$, \,\,\,\, $\gamma\!=\!\gamma_0/4$.]
    {
    \includegraphics[width=0.46\textwidth]{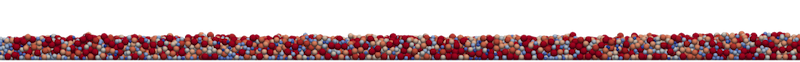}
    \label{fig:sv_10}
   }
    \subfigure[$t_0\!=\!3d_{max,0}$, \,\,\,\, $\gamma\!=\!\gamma_0$.]
       {
    \includegraphics[width=0.46\textwidth]{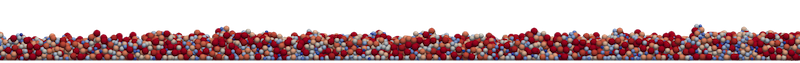}
    \label{fig:sv_11}
   }
    \hspace{0.1 cm}
    \subfigure[$t_0\!=\!3d_{max,0}$, \,\,\,\, $\gamma\!=\!4\gamma_0$.]
       {
    \includegraphics[width=0.46\textwidth]{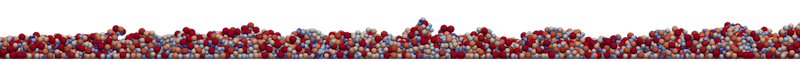}
    \label{fig:sv_12}
   }
    \subfigure[$t_0\!=\!4d_{max,0}$, \,\,\,\, $\gamma\!=\!0$.]
       {
    \includegraphics[width=0.46\textwidth]{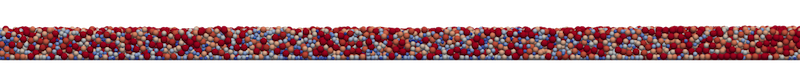}
    \label{fig:sv_13}
   }
    \hspace{0.1 cm}
    \subfigure[$t_0\!=\!4d_{max,0}$, \,\,\,\, $\gamma\!=\!\gamma_0/4$.]
       {
    \includegraphics[width=0.46\textwidth]{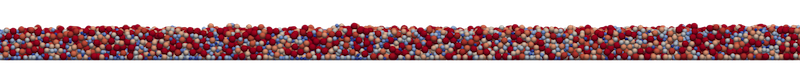}
    \label{fig:sv_14}
   }
    \subfigure[$t_0\!=\!4d_{max,0}$, \,\,\,\, $\gamma\!=\!\gamma_0$.]
       {
    \includegraphics[width=0.46\textwidth]{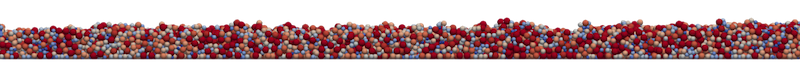}
    \label{fig:sv_15}
   }
    \hspace{0.1 cm}
    \subfigure[$t_0\!=\!4d_{max,0}$, \,\,\,\, $\gamma\!=\!4\gamma_0$.]
       {
    \includegraphics[width=0.46\textwidth]{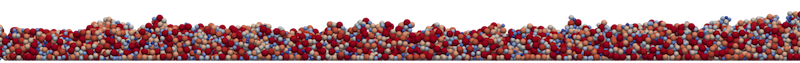}
    \label{fig:sv_16}
   }
   \caption{Resulting powder layers for different layer thicknesses $t_0$ and surface energies $\gamma$: Side view on powder bed.}
  \label{fig:sv}
\end{figure}

In this section, the influence of the powder layer thickness will be investigated. Figures~\ref{fig:tv} and~\ref{fig:sv} illustrate the (top and side views of the) final powder layer configurations resulting from the nominal powder layer thicknesses $t_0=d_{max,0},2d_{max,0},3d_{max,0},4d_{max,0}$ (with nominal maximal powder particle diameter $d_{max,0}=50\mu m$) as well as from the surface energy values $\gamma=0,\gamma_0/4,\gamma_0,4\gamma_0$.\\

According to Figure~\ref{fig:tv}, a low powder layer thickness $t_0=d_{max,0}$ in the range of the maximal particle diameter leads to rather sparse particle distributions, but not to continuous and homogeneous powder layers. Similar observations have e.g. been made by~\cite{Mindt2016}. Only at the right end of the powder bed, where the powder bed edge (not illustrated) induces shear forces onto the bulk powder, a slightly increased packing density can be observed (at least for the less cohesive powders). With increasing nominal thickness $t_0$, the powder layer becomes increasingly continuous. For the less cohesive powders $\gamma=0$ and $\gamma=\gamma_0/4$ this state is already reached at $t_0=2d_{max,0}$, while the more cohesive powders $\gamma=\gamma_0$ and $\gamma=4\gamma_0$ require a layer thickness of $t_0=3d_{max,0}$ and $t_0=4d_{max,0}$, respectively, in order to end up with a continuous powder layer. It is well-known that increasingly cohesive powders are characterized by lower packing fractions. Moreover, the particle agglomerates within highly cohesive powders are subject to higher and stronger varying resistance forces when passing through the gap of the recoating blade, which, in turn, results in less dense and rather irregular powder layers in the range of small blade gaps. Larger agglomerates that do not fit through this gap are sheared off as long as the adhesion between the bulk powder and the substrate is high enough to guarantee for stick friction. Eventually, Figure~\ref{fig:sv} visualizes how the surface roughness increases with increasing cohesion / surface energy, which can partly be attributed to the presence of larger particle agglomerates, but also to particles that are ripped out of the layer compound due to particle-to-blade adhesion. It has to be mentioned that the 2D projections employed in Figure~\ref{fig:sv} lead to an averaging effect in (the periodic) y-direction and, thus, to 1D surface profiles that look smoother than the (actual) 2D surfaces are. This effect can easily be observed when comparing the 2D projections in Figures~\ref{fig:sv_9} and~\ref{fig:sv_11} with the corresponding 3D plots in Figures~\ref{fig:metrics3_4} and~\ref{fig:metrics3_3}, derived from the same numerical simulations. A similar "artificial smoothing effect" also occurs in approaches that apply such a projected 1D surface profile for the statistical evaluation of the surface profile / roughness~\citep{Parteli2016}. In order to avoid this undesirably smoothing of the surface roughness, the real 2D surface profiles (see Figure~\ref{fig:metrics3}) are employed for statistical evaluation throughout this work.\\

\begin{figure}[h!!]
   \centering
   \subfigure[Mean value of packing fraction field $\Phi_t(x,y)$.]
    {
    \includegraphics[width=0.48\textwidth]{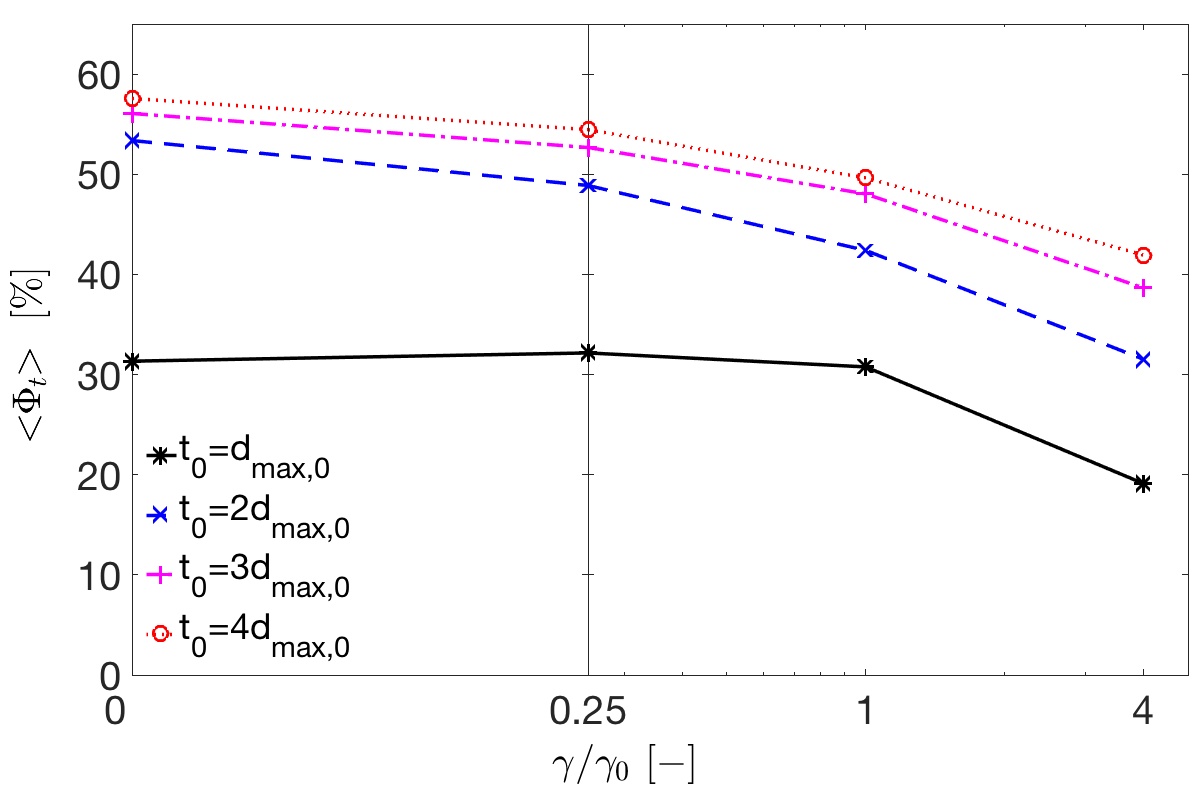}
    \label{fig:thickness_1}
   }
  \hspace{0.1 cm}
   \centering
 \subfigure[Mean value of surface profile field $z_{int}(x,y)$.]
   {
    \includegraphics[width=0.48\textwidth]{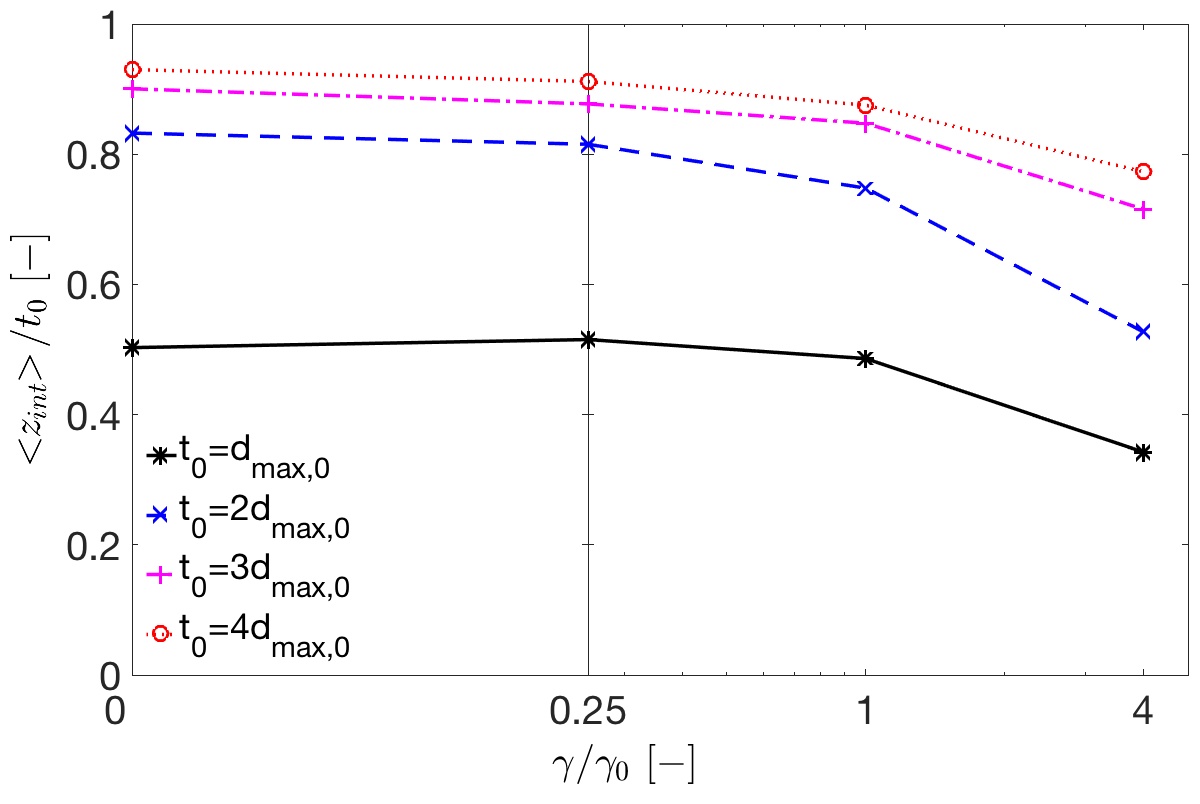}
    \label{fig:thickness_2}
   }    
   \centering
    \subfigure[Standard deviation of packing fraction field $\Phi_t(x,y)$.]
    {
    \includegraphics[width=0.48\textwidth]{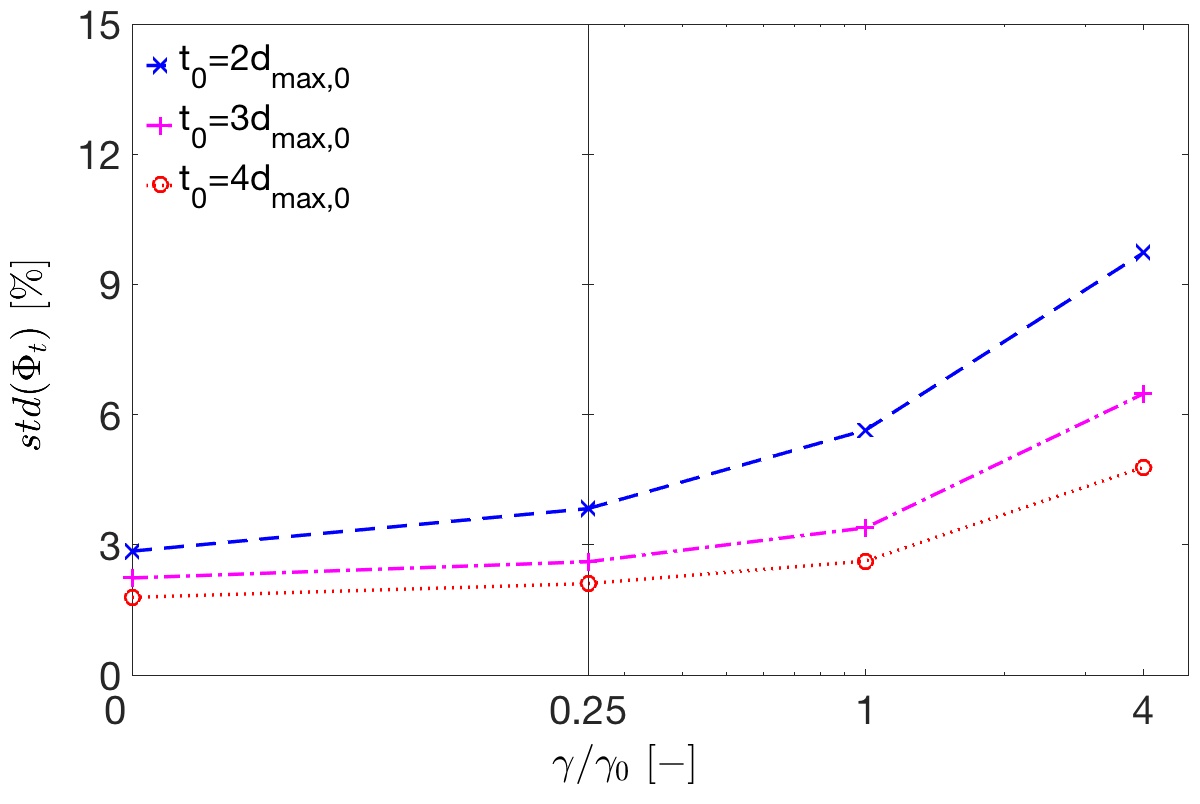}
    \label{fig:thickness_3}
   }
   \hspace{0.1 cm}
 \centering
 \subfigure[Standard deviation of surface profile field $z_{int}(x,y)$.]
   {
    \includegraphics[width=0.48\textwidth]{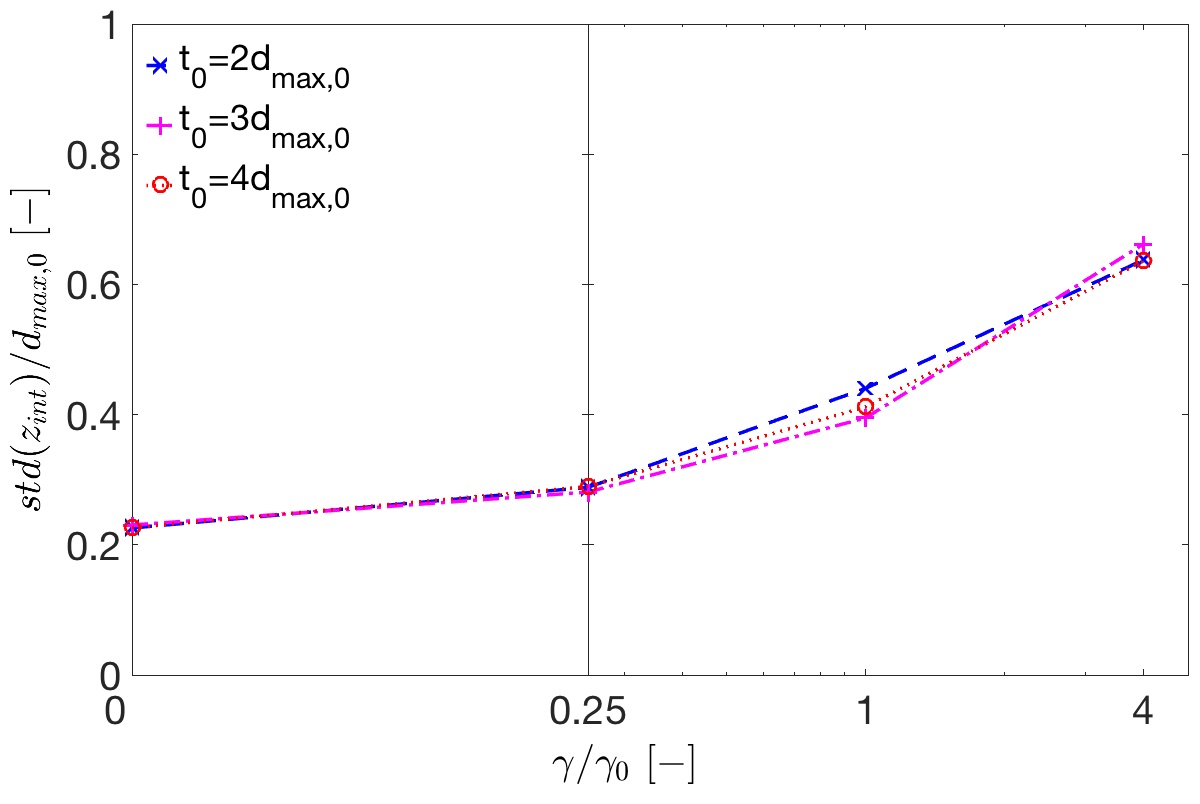}
    \label{fig:thickness_4}
   }
   \caption{Mean value and standard deviation of $\Phi_t(x,y)$ and $z_{int}(x,y)$ as function of powder cohesiveness $\gamma / \gamma_0$ for different nominal layer thicknesses $t_0$.}
  \label{fig:thickness}
\end{figure}

Figure~\ref{fig:thickness_1} illustrates the mean packing fraction $<\!\Phi_t(x,y)\!>$ for the different powder layer thicknesses. It is well-known that the packing fraction in cohesive powders decreases with increasing adhesion forces between the particles, which hinders the gravity-driven settling of the particles~\citep{Walton2008,Yang2000}. This effect can also be observed in Figure~\ref{fig:thickness_1}. Moreover, the packing densities increase with increasing layer thickness since the particles can increasingly arrange in an optimal manner and the relative influence of (negative) boundary effects at the bottom and the top of the powder layer decreases. The difference between the cases $t_0=3d_{max,0}$ and $t_0=4d_{max,0}$ is already very small, and, thus, the layer thickness $t_0=3d_{max,0}$ can already be considered as reasonably close to the optimum. Obviously, the extreme case $t_0=d_{max,0}$ does not lead to a continuous powder layer and, thus, results in a very low packing density. This discontinuous powder layer leads to non-representative standard deviations of the fields $\Phi_t(x,y)$ and $z_{int}(x,y)$. Thus, for reasons of better comprehensibility, this case will not be considered in the subsequent standard deviation plots.\\

According to Figure~\ref{fig:thickness_3}, the associated standard deviation $std(\Phi_t(x,y))$ of the packing fraction field increases with increasing powder cohesiveness. This behavior might be explained by the following two arguments: First, with increasing cohesiveness, particle agglomerates become larger compared to the bin size underlying the packing fraction calculation, which results in less packing fraction averaging across individual bins. Second, surface roughness increases with increasing adhesion (see below), and these surface profile variations between minimal and mean layer height (chosen as upper bound of the bins) also lead to packing fraction variations. The decreasing influence of these surface effects with increasing layer thickness is one reason, why also the packing fraction variations decrease with increasing layer thickness. A second reason why packing fraction variations decrease with increasing layer thickness - and why this effect is more pronounced in the regime of high surface energies - is the increasing ratio of bin height / volume to the size of powder particles / agglomerates when increasing the layer thickness, which again leads to an increased averaging effect.\\

\begin{figure}[h!!]
   \centering
   \subfigure[Alternative definitions of mean packing fraction ($t_0=4d_{max,0}$).]
    {
    \includegraphics[width=0.48\textwidth]{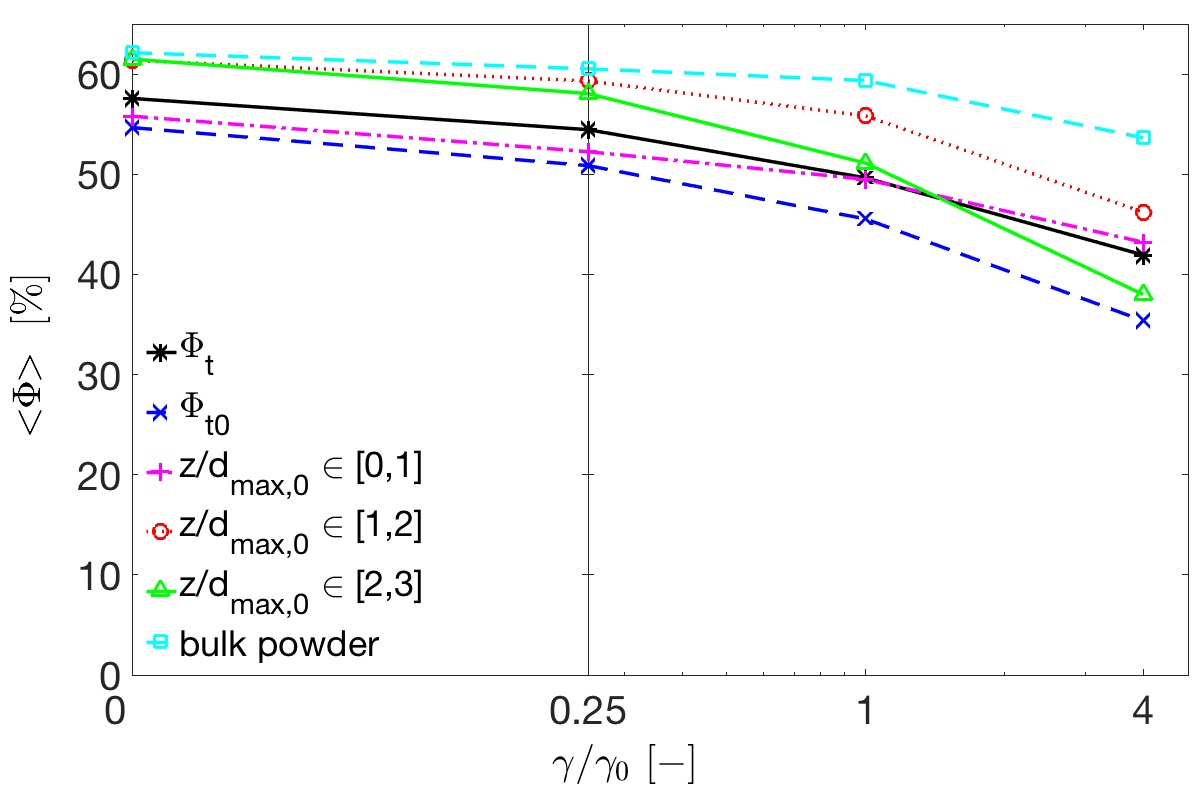}
    \label{fig:thicknessb_1}
   }
  \hspace{0.1 cm}
   \centering
 \subfigure[Relative error in mean surface profile height $t:=<z_{int}(x,y)>$.]
   {
    \includegraphics[width=0.48\textwidth]{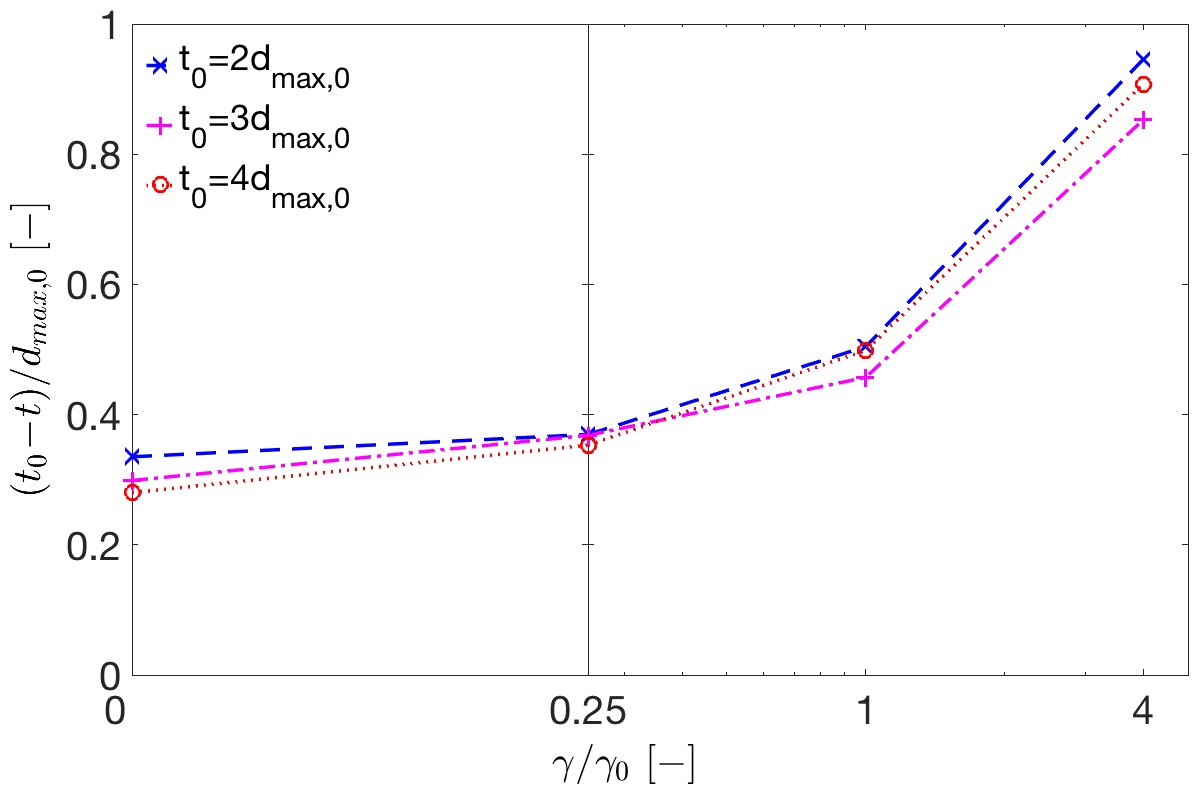}
    \label{fig:thicknessb_2}
   }    
   \caption{Alternative metrics for evaluation of mean packing fraction and surface profile height for different nominal layer thicknesses $t_0$.}
  \label{fig:thicknessb}
\end{figure}

Figures~\ref{fig:thickness_2} and~\ref{fig:thickness_4} visualize the mean surface profile height $t:=<\!z_{int}(x,y)\!>$ as well as the associated standard deviation $std(z_{int}(x,y))$, i.e. the surface roughness. Accordingly, the surface roughness increases with the cohesiveness of the powder. This observation can partly be attributed to the presence of larger particle agglomerates, but also to particles that are ripped out of the layer compound near the surface due to particle-to-blade adhesion. Since the height of surface asperities is naturally bounded by the lower edge of the recoating blade, this higher surface roughness in turn leads to a decreasing mean value $t$ of the surface profile when increasing the surface energy (see Figures~\ref{fig:thickness_2}). According to Figure~\ref{fig:thickness_2}, the ratio of mean layer height $t$ to nominal layer height $t_0$ increases with increasing nominal layer height between $t_0=2d_{max,0}$ and $t_0=4d_{max,0}$. However, this effect can mainly be attributed to the increasing factor $t_0$ used for normalization. For comparison, Figure~\ref{fig:thicknessb_2} represents the relative error $(t-t_0)/d_{max,0}$ between real and nominal layer height normalized with the constant factor $d_{max,0}$, and suggests that this error is almost independent of the nominal layer thickness $t_0$. The same observation can be made for the standard deviation of the surface profile in Figure ~\ref{fig:thickness_4}. It is interesting to realize that for layer thicknesses of $t_0=2d_{max,0}$ and above only the relative deviations between the real and nominal surface profile depend on the nominal layer thickness $t_0$, but not the absolute deviations. In other words:  Even though the nominal thicknesses $t_0=3d_{max,0}$ and $t_0=4d_{max,0}$ lead to (comparatively) continuous powder layers with only a few blank spots on the substrate, these powder layers comprise a top region of height $2d_{max,0}$ with similar characteristics as the discontinuous powder layer resulting from the choice $t_0=2d_{max,0}$.\\

At the end of this section, alternative definitions of the mean packing fraction $<\!\Phi(x,y)\!>$ based on different representative volumes and the case $t_0=4d_{max,0}$ (see Figure~\ref{fig:thicknessb_1}) shall be discussed and compared to the results presented above. As expected, the alternative definition $<\!\Phi_{t0}(x,y)\!>$ considering the entire nominal powder layer volume of height $t_0$ leads to a lower mean packing fraction than the variant $<\!\Phi_{t}(x,y)\!>$ limited by the real mean layer height $t$. With increasing relative deviation between the real and nominal layer thicknesses $t$ and $t_0$ (e.g. in the range of small nominal layer thicknesses $t_0$ or high blade velocities $V_B$, see Section~\ref{sec:recoating_bladevelocity}), the alternative (and frequently employed) metric $<\!\Phi_{t0}(x,y)\!>$ increasingly underestimates the actual packing fraction values within the powder layer, and is consequently less suitable for powder layer characterization. In addition to the two metrics $<\!\Phi_{t}(x,y)\!>$ and $<\!\Phi_{t0}(x,y)\!>$, also mean packing fractions calculated for powder sub-layers limited by the z-coordinate intervals $z \in [0; d_{max,0}]$ (bottom layer close to the substrate), $z \in [d_{max,0};2d_{max,0}]$ and $z \in [2d_{max,0};3d_{max,0}]$ are presented in Figure~\ref{fig:thicknessb_1} (even for the case $\gamma=4\gamma_0$, the z-coordinate $z=3d_{max,0}$ is below the coordinate associated with the mean layer height $<\!z_{int}\!>|_{\gamma = 4\gamma_0} \approx 3.1 d_{max,0}$, see Figure~\ref{fig:thicknessb_2}). For the non-cohesive case $\gamma=0$, the aforementioned boundary effect of reduced packing fraction at the bottom of the powder layer can observed: While the upper layers $z \in [d_{max,0};2d_{max,0}]$ and $z \in [2d_{max,0};3d_{max,0}]$ have almost identical packing fractions (of approximately 62 \%) above the mean value $<\!\Phi_{t}(x,y)\!>$ (approximately 58 \%), the bottom layer $z \in [0; d_{max,0}]$ shows a reduced packing fraction (of approximately 56 \%). Obviously, the surface roughness in the top layer $z \in [3d_{max,0};t]$ (not shown in Figure~\ref{fig:thicknessb_2}) must lead to an even smaller packing fraction to end up with the mean value $<\!\Phi_{t}(x,y)\!>$ across the entire layer thickness $z \in [0;t]$. With increasing powder cohesiveness, this behavior changes: For the case $\gamma=4\gamma_0$, powder layer depressions due to surface roughness reach deep into the layer $z \in [2d_{max,0};3d_{max,0}]$ (third from the bottom) resulting in lower packing fractions than in the very bottom layer $z \in [0; d_{max,0}]$. This observation can easily be verified by subtracting the standard deviation of the surface profile height ($std(z_{int})|_{\gamma = 4\gamma_0} \approx 0.6 d_{max,0}$, Figure~\ref{fig:thickness_4}) from the corresponding mean value ($<\!z_{int}\!>|_{\gamma = 4\gamma_0}\approx 3.1 d_{max,0}$, see Figure~\ref{fig:thicknessb_2}), which yields a value of $z_{int}=2.5 d_{max,0} \in [2d_{max,0};3d_{max,0}]$. All in all, the sub-layer $z \in [d_{max,0};2d_{max,0}]$ is least biased by boundary effects on the top and bottom of the powder bed, and, thus, the resulting packing fraction is closest to packing fraction values measured for larger bulk powder volumes.\\

For comparison, the packing fraction has also been determined for the bulk powder in the powder reservoir after the initial, gravity-driven settling (see e.g. Figure~\ref{fig:model_configs1}). The corresponding results of the four cases $\gamma=0, \, \gamma_0/4, \, \gamma_0, \, 4\gamma_0$ are plotted in Figure~\ref{fig:thicknessb_1} as well. Accordingly, mean packing fraction values of approximately $<\!\Phi\!>|_{\gamma = 0}=62\%$ and $<\!\Phi\!>|_{\gamma = 4\gamma_0}=54\%$ result for the initial powder configuration and the surface energies $\gamma=0$ and $\gamma=4\gamma_0$, respectively, while the corresponding values within the (densest) sub-layer $z \in [d_{max,0};2d_{max,0}]$ of the recoated powder are $<\!\Phi\!>|_{\gamma = 0}=61\%$ and $<\!\Phi\!>|_{\gamma = 4\gamma_0}=46\%$ (see Figure~\ref{fig:thicknessb_1}). Two conclusions can be drawn from these deviations: First, the packing fractions determined for the bulk powder after gravity-driven settling are still higher than the (maximal) values prevalent in the recoated powder layer. For the case $\gamma=4\gamma_0$, this difference is considerable (reduction from $54\%$ to $46\%$). This might be explained by remaining surface roughness effects that still influence the resulting packing fraction, even for the sub-layer $z \in [d_{max,0};2d_{max,0}]$ with maximal packing fraction. Moreover, also a reduced level of compression forces (as compared to the gravity-driven powder settling) as well as the presence of shear and tension forces during the recoating of cohesive powders can explain this loosening of the powder packing during recoating. Second, the drop of packing fraction from $62\%$ to $54\%$ when increasing the adhesion-to-gravity force ratio $F_{\gamma}/F_G$ from $0$ to approximately $50$ (which represents the case $\gamma=4\gamma_0$) is in very good agreement with corresponding results by~\cite{Yang2000} (see Figure 4 of this reference evaluated for a particle diameter of $d=65 \mu m$, which yields a ratio $F_{\gamma}/F_G \approx 50$), despite the slightly different system parameters chosen in this reference (i.e. smaller friction coefficient, uni-sized powder instead of size distribution). This result serves as further verification of the present modeling approach and the general strategy of analyzing dimensionless adhesion-to-gravity force ratios. Moreover, this conformity of results also suggests that the differing parameters of the two approaches (friction coefficient, specific particle size distribution) play only a secondary role compared to the adhesion-to-gravity force ratio (determined on the basis of the mean particle size).\\

Finally, it shall again be emphasized that the packing fraction values determined for thin powder layers might be strongly influenced by boundary effects on the bottom and the top of the powder layer. For example, only slight changes in the representative volume considered for packing fraction calculation can readily lead to drastically different results. These effects need to be taken into account when comparing different packing fraction values stated in the literature. Throughout this work, the packing fraction $<\!\Phi_t(x,y)\!>$ defined across the mean layer height $z \in [0;t]$ will be applied. On the one hand, this definition avoids deterioration of packing fraction results for powder layers of reduced thickness $t$ (see e.g. Figure~\ref{fig:velocityb_3}) as it would be the case for the variant $<\!\Phi_{t0}(x,y)\!>$. On the other hand, it does include boundary effects on the bottom and top of the powder layer, which are typical for the thin powder layers employed in metal AM and relevant for the subsequent melting process.\\

\subsection{Influence of blade velocity}
\label{sec:recoating_bladevelocity}

\begin{figure}[h!!]
   \centering
   \subfigure[Mean value of packing fraction field $\Phi_t(x,y)$.]
    {
    \includegraphics[width=0.48\textwidth]{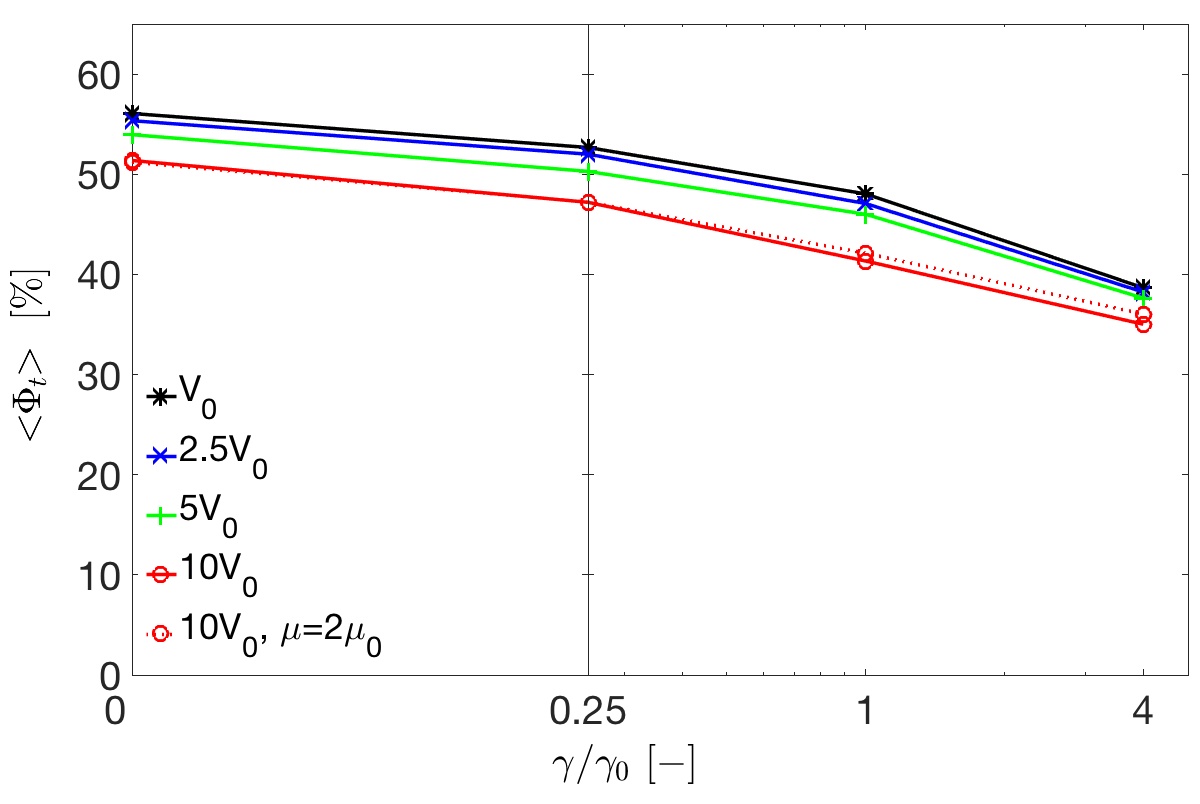}
    \label{fig:velocity_1}
   }
  \hspace{0.1 cm}
   \centering
 \subfigure[Mean value of surface profile field $z_{int}(x,y)$.]
   {
    \includegraphics[width=0.48\textwidth]{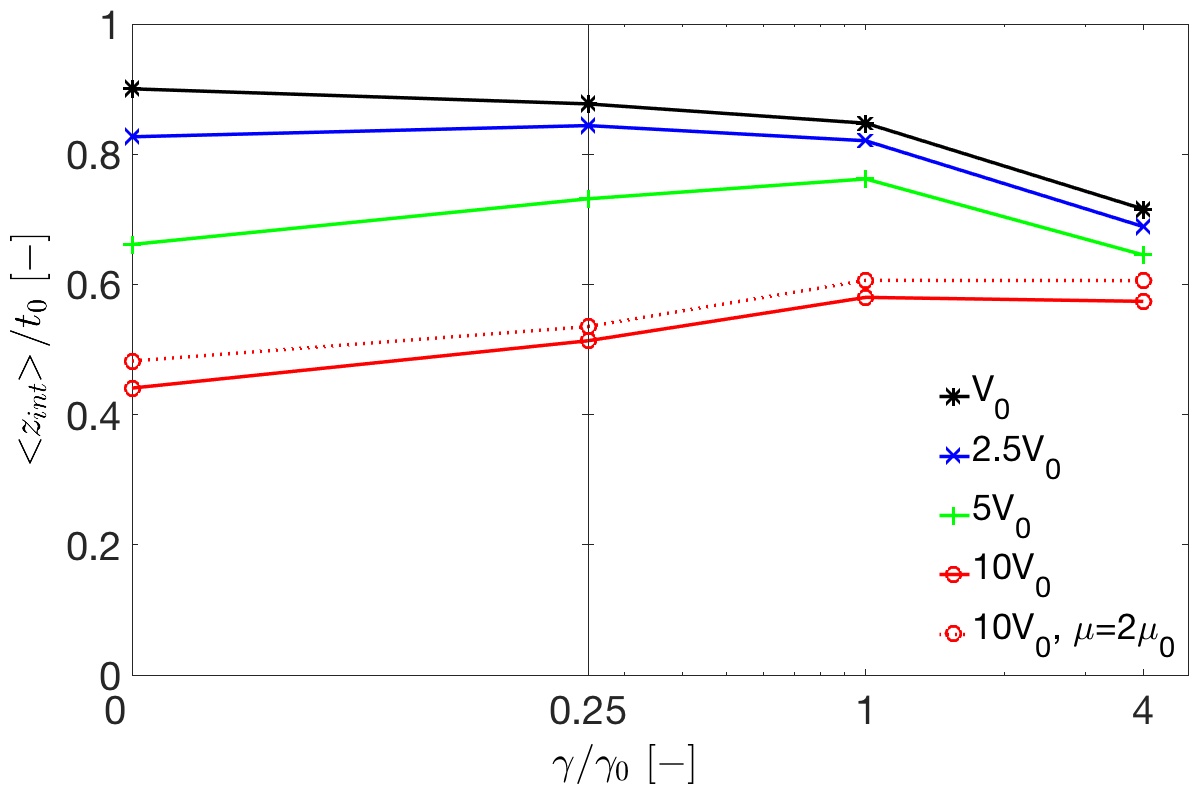}
    \label{fig:velocity_2}
   }    
   \centering
    \subfigure[Standard deviation of packing fraction field $\Phi_t(x,y)$.]
    {
    \includegraphics[width=0.48\textwidth]{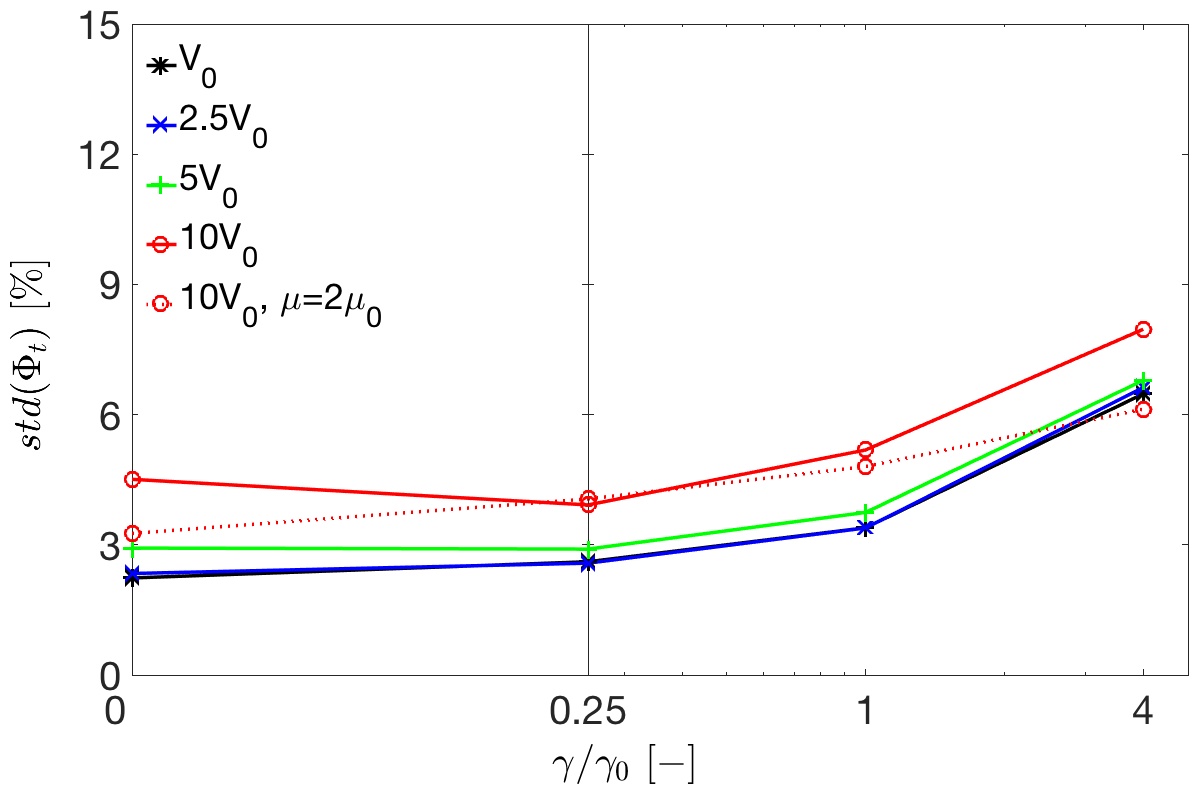}
    \label{fig:velocity_3}
   }
   \hspace{0.1 cm}
 \centering
 \subfigure[Standard deviation of surface profile field $z_{int}(x,y)$.]
   {
    \includegraphics[width=0.48\textwidth]{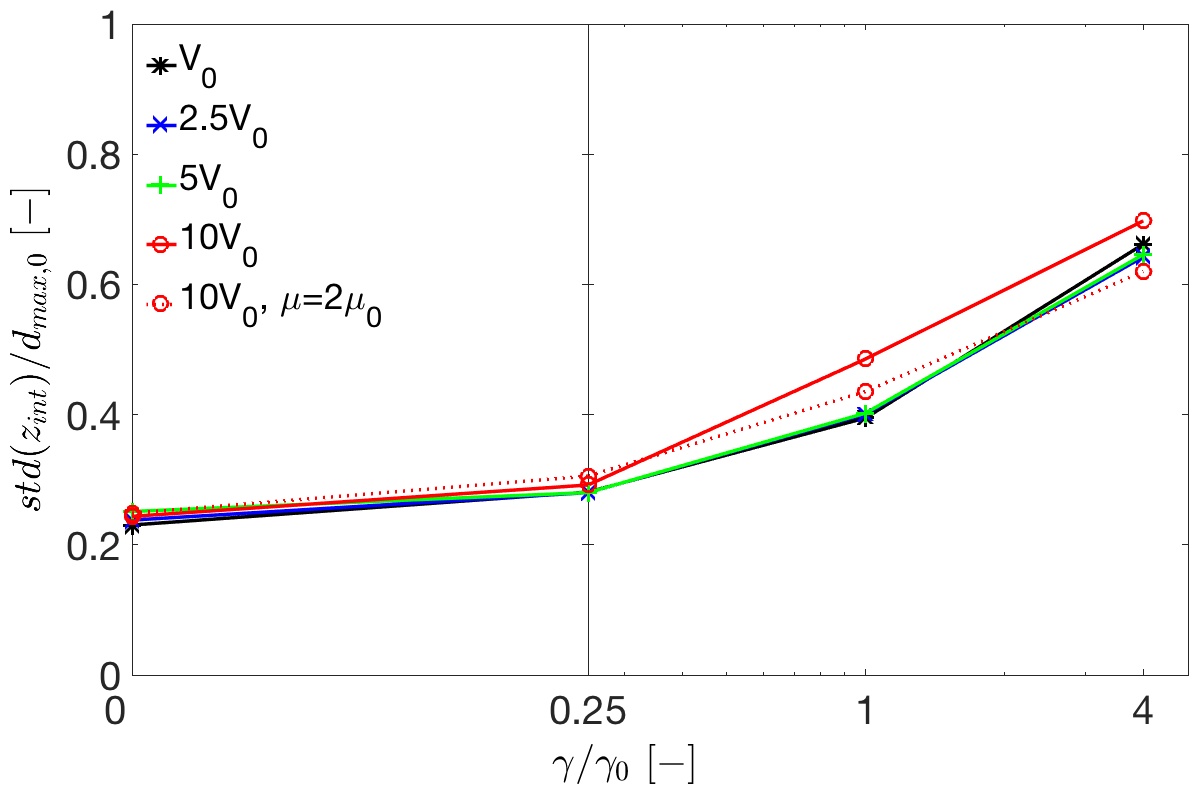}
    \label{fig:velocity_4}
   }
   \caption{Mean value and standard deviation of $\Phi_t(x,y)$ and $z_{int}(x,y)$ as function of powder cohesiveness $\gamma / \gamma_0$ for different blade velocities $V_B$ ($t_0=3d_{max,0}$).}
  \label{fig:velocity}
\end{figure}

While so far only the ideal case of quasi-static powder spreading has been considered, this section will focus on the influence of spreading velocity. It is known from existing experimental~\citep{Meyer2017} and numerical~\citep{Mindt2016,Parteli2016} studies that increased blade velocity typically deteriorates the powder layer quality. Here, this deterioration will be quantified by means of the proposed metrics as illustrated in Figure~\ref{fig:velocity}. Thereto, the four different blade velocities $V_B=V_0$, $2.5V_0$, $5V_0$ and $10V_0$ will be investigated. A special focus will lie on the influence of adhesion.\\

\begin{figure}[h!!]
   \centering
    \subfigure[Intermediate configuration: $\gamma\!=\!0$.]
   {
    \includegraphics[width=0.46\textwidth]{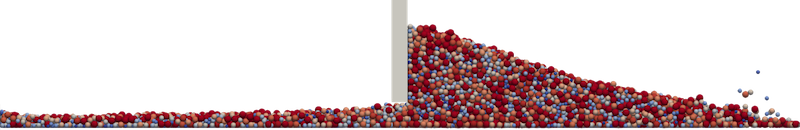}
    \label{fig:velocityb_1}
   }
    \hspace{0.1 cm}
    \subfigure[Intermediate configuration: $\gamma\!=\!\gamma_0$.]
    {
    \includegraphics[width=0.46\textwidth]{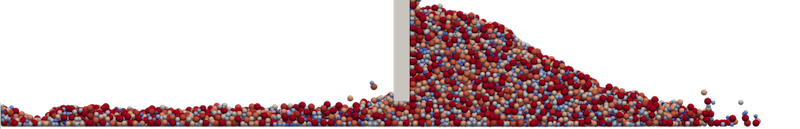}
    \label{fig:velocityb_2}
   }
   \subfigure[Final configuration: $\gamma\!=\!0$.]
   {
    \includegraphics[width=0.46\textwidth]{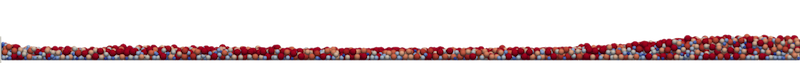}
    \label{fig:velocityb_3}
   }
    \hspace{0.1 cm}
    \subfigure[Final configuration: $\gamma\!=\!\gamma_0$.]
    {
    \includegraphics[width=0.46\textwidth]{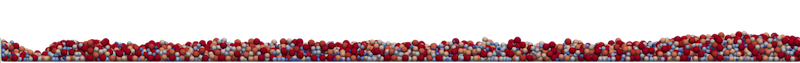}
    \label{fig:velocityb_4}
   }
   \caption{Intermediate and final recoating configuration for increased blade velocity $V_B=10V_0$ and the cases $\gamma=0$ and $\gamma=\gamma_0$ ($t_0=3d_{max,0}$).}
  \label{fig:velocityb}
\end{figure}

Among the four metrics in Figure~\ref{fig:velocity}, the highest sensitivity with respect to blade velocity can be observed for the mean layer height (Figure~\ref{fig:velocity_2}). It is characterized by two effects: First, the mean layer height $t\!=\!<\!z_{int}(x,y)\!>$ decreases drastically with increasing blade velocity. For $V_B=10V_0$ and $\gamma=0$, the mean layer height is less than half of the nominal layer height $t_0$. Second, the effect of increased blade velocity decreases with increasing adhesion. For blade velocities $V_B \geq 5V_0$, the mean layer height even increases when increasing the adhesion from $\gamma=0$ to $\gamma=\gamma_0$. This observation shall be explained by means of Figure~\ref{fig:velocityb} visualizing an intermediate and the final configuration of the recoating process for the cases $\gamma=0$ and $\gamma=\gamma_0$. By comparing Figures~\ref{fig:velocityb_1} and~\ref{fig:velocityb_3}, it becomes obvious that the layer thickness of the non-cohesive powder further decreases with time at local positions that have already been passed by the recoating blade. This means that the momentum induced by the recoating blade is high enough to sustain the powder flow for some time even though the blade has already passed the considered location. Increasing adhesion reduces the flowability of the powder, i.e. the powder sticks together and to the substrate, which in turn reduces the post-flow of powder as well as the resulting layer thickness reduction. While it is typically claimed that powder of high flowability is beneficial for the power spreading process in metal additive manufacturing, the present results suggest that actually a reduced flowability (e.g. through increased inter-particle friction and/or adhesion) of the powder might even improve certain quality metrics in the range of increased blade velocities. In this context, it has to be noted that blade velocities in the range of $V_B=10V_0=100mm/s$ and above are typical for practical AM processes. In order to underline the above statement, a further simulation with high blade velocity $V_B=10V_0$ and increased friction $\mu=2\mu_0$ has been conducted and visualized in Figure~\ref{fig:velocity}. As expected, the higher friction coefficient reduces the flowability of the powder, which in turn results in a slightly increased layer height.\\

In the highly cohesive case $\gamma=4\gamma_0$, the negative effects of adhesion (see Section~\ref{sec:recoating_thickness}) outweigh the gain due to reduced flowability, and the mean layer thicknesses decreases again from $\gamma=\gamma_0$ to $\gamma=4\gamma_0$, even in the range of higher blade velocities. However, the layer height difference between the cases $V_B=V_0$ and $V_B=10V_0$ continuously decreases from $\gamma=0$ to $\gamma=4\gamma_0$. Eventually, the influence of the blade velocity on the other metrics in Figures~\ref{fig:velocity_1},~\ref{fig:velocity_3} and~\ref{fig:velocity_4} might be explained by the lower layer thicknesses resulting from the higher blade velocities. A comparison with Figure~\ref{fig:thickness} largely confirms this hypothesis. The lower velocity-dependence of these metrics in the range of high surface energies correlates well with the lower velocity-dependence of the layer thickness in the range of high surface energies as discussed above (see Figure~\ref{fig:velocity_2}). In contrast to the mean layer thickness $t$, however, the mean packing density $<\!\Phi_t(x,y)\!>$ still decreases with increasing adhesion, even for the case $V_B=10V_0$. Thus, the negative effect of increased velocity on this metric can not be compensated by reduced flowability.\\

It can be concluded that the net effect of increased blade velocities is comparable to decreased layer thicknesses. Counterintuitively, however, lower flowability of the powder might - at least partly - reduce the negative influence of increased blade velocity. Finally, a remark shall be made concerning scaleability of these results: For all quasi-static investigations throughout this work, it has been stated that an increase / decrease of surface energy by a factor of four is equivalent to a decrease / increase of the mean particle diameter by a factor of two. The situation changes slightly in the regime of higher blade velocities, when inertia effects begin to play a role. In principle, the results in Figure~\ref{fig:velocity} are still valid if mean particle size is varied instead of surface energy. However, in the regime of higher blade velocities, the blade velocities would have to be consistently down-scaled with the particle size in order to end up with results that are equivalent to the ones presented here.\\

\subsection{Influence of blade and substrate properties}
\label{sec:recoating_bladeandsubstrate}

For simplicity, it has been assumed so far that the parameters controlling the particle-to-particle interaction as well as the particle-to-wall interaction (e.g. adhesion, friction, coefficient of restitution) are identical. In practical metal AM processes, this is rather unlikely, however, which might have considerable (wanted or unwanted) consequences on the result of the powder spreading process. Again, the notation particle-to-wall interaction includes all interactions of particles with surrounding solid components and walls. Specifically, it includes the interaction between particles and recoating blade (particle-to-blade interaction) as well as the interaction between particles and the bottom / side walls of the powder bed (particle-to-substrate interaction).This section will especially focus on the influence of particle-to-blade and particle-to-substrate adhesion.\\

\begin{figure}[h!!]
   \centering
    \subfigure[Side view: $\gamma_B=\gamma$.]
   {
    \includegraphics[width=0.46\textwidth]{Figs/recoating_21000_5x1_adhesion_4e7_new_sv.png}
    \label{fig:nba_1}
   }
    \hspace{0.1 cm}
    \subfigure[Side view: $\gamma_B=0$.]
    {
    \includegraphics[width=0.46\textwidth]{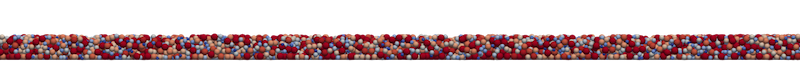}
    \label{fig:nba_2}
   }
   \subfigure[Top view: $\gamma_B=\gamma$.]
   {
    \includegraphics[width=0.46\textwidth]{Figs/recoating_21000_5x1_adhesion_4e7_new_tv.png}
    \label{fig:nba_3}
   }
    \hspace{0.1 cm}
    \subfigure[Top view: $\gamma_B=0$.]
    {
    \includegraphics[width=0.46\textwidth]{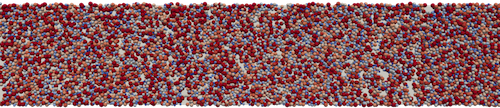}
    \label{fig:nba_4}
   }
   \caption{Resulting powder layer in dependence of the particle-to-blade adhesion $\gamma_B$ for the case $\gamma=4\gamma_0$ ($t_0=3d_{max,0}$).}
  \label{fig:nba}
\end{figure}

\begin{figure}[t!!]
   \centering
   \subfigure[Mean value of packing fraction field $\Phi_t(x,y)$.]
    {
    \includegraphics[width=0.48\textwidth]{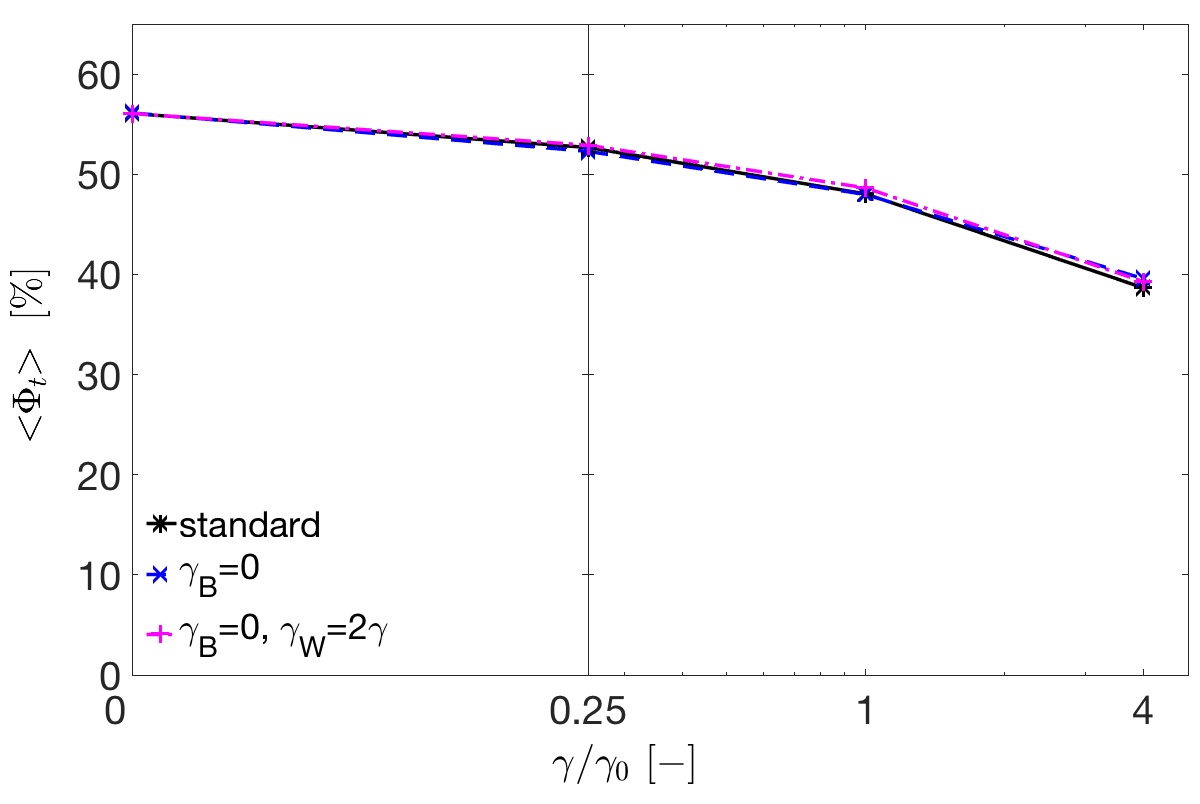}
    \label{fig:wall_1}
   }
  \hspace{0.1 cm}
   \centering
 \subfigure[Mean value of surface profile field $z_{int}(x,y)$.]
   {
    \includegraphics[width=0.48\textwidth]{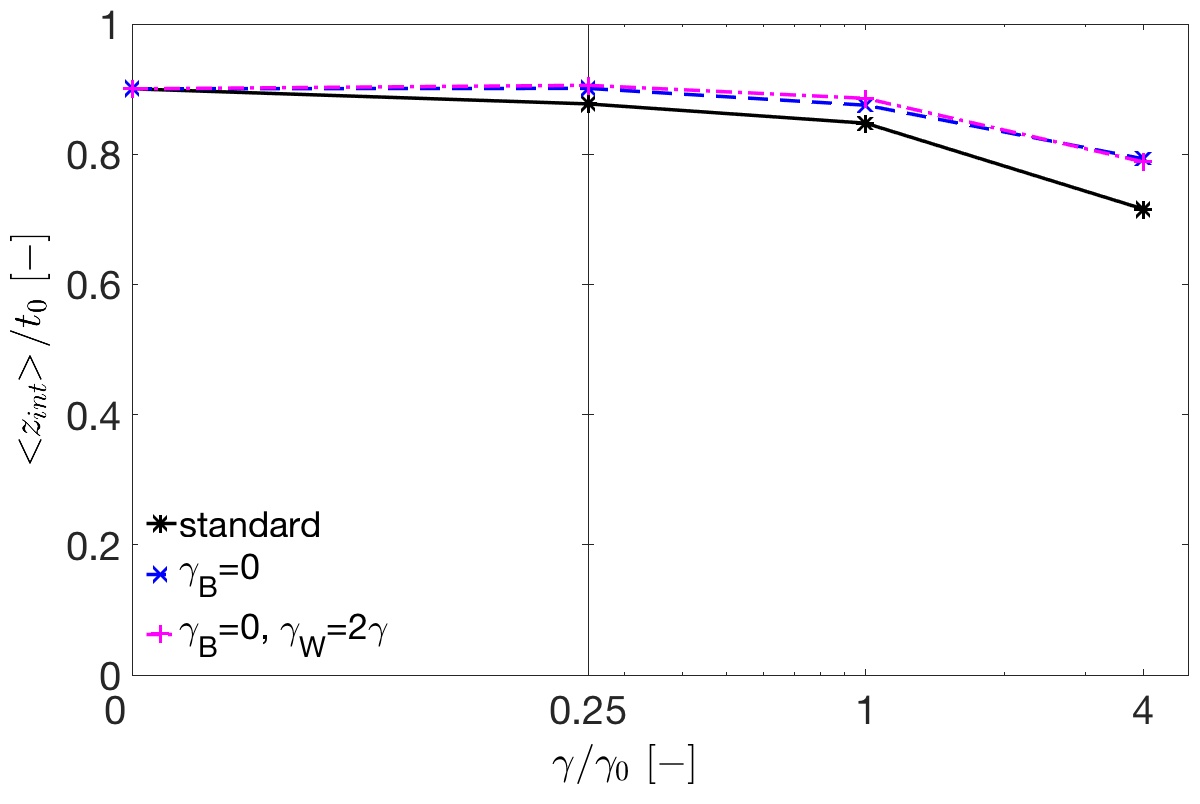}
    \label{fig:wall_2}
   }    
   \centering
    \subfigure[Standard deviation of packing fraction field $\Phi_t(x,y)$.]
    {
    \includegraphics[width=0.48\textwidth]{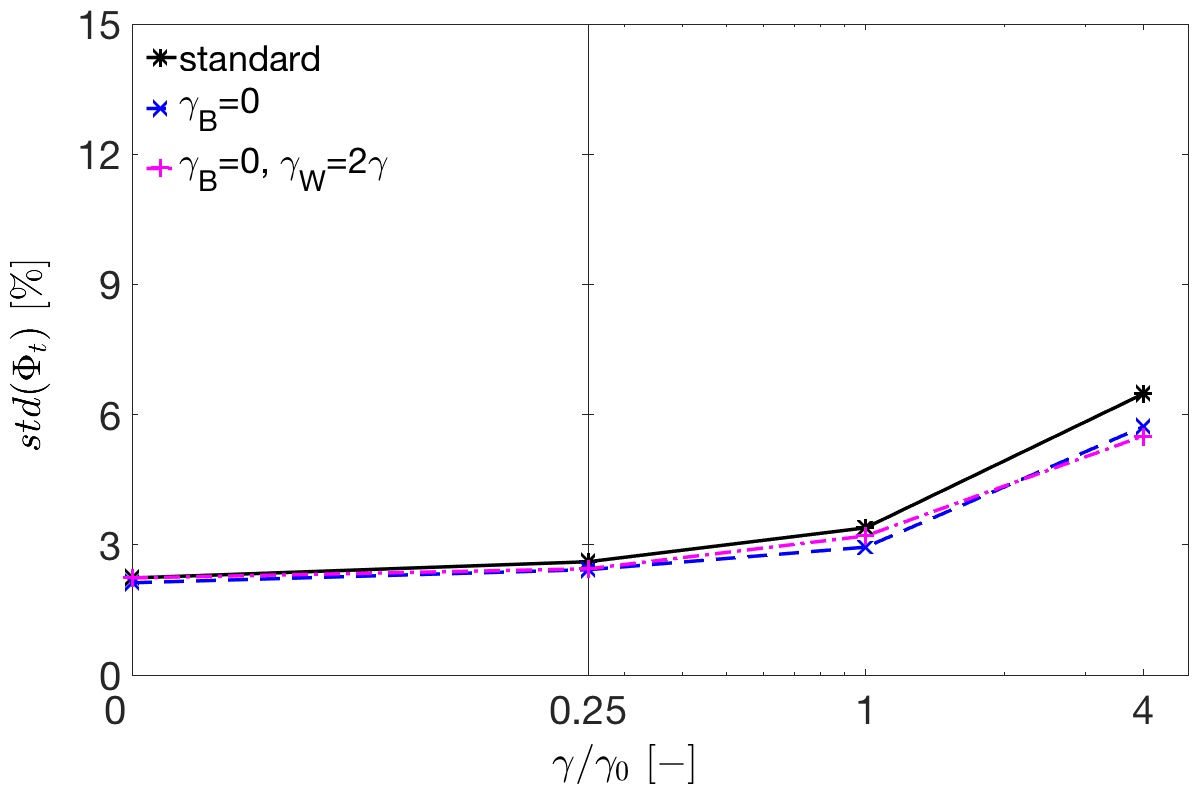}
    \label{fig:wall_3}
   }
   \hspace{0.1 cm}
 \centering
 \subfigure[Standard deviation of surface profile field $z_{int}(x,y)$.]
   {
    \includegraphics[width=0.48\textwidth]{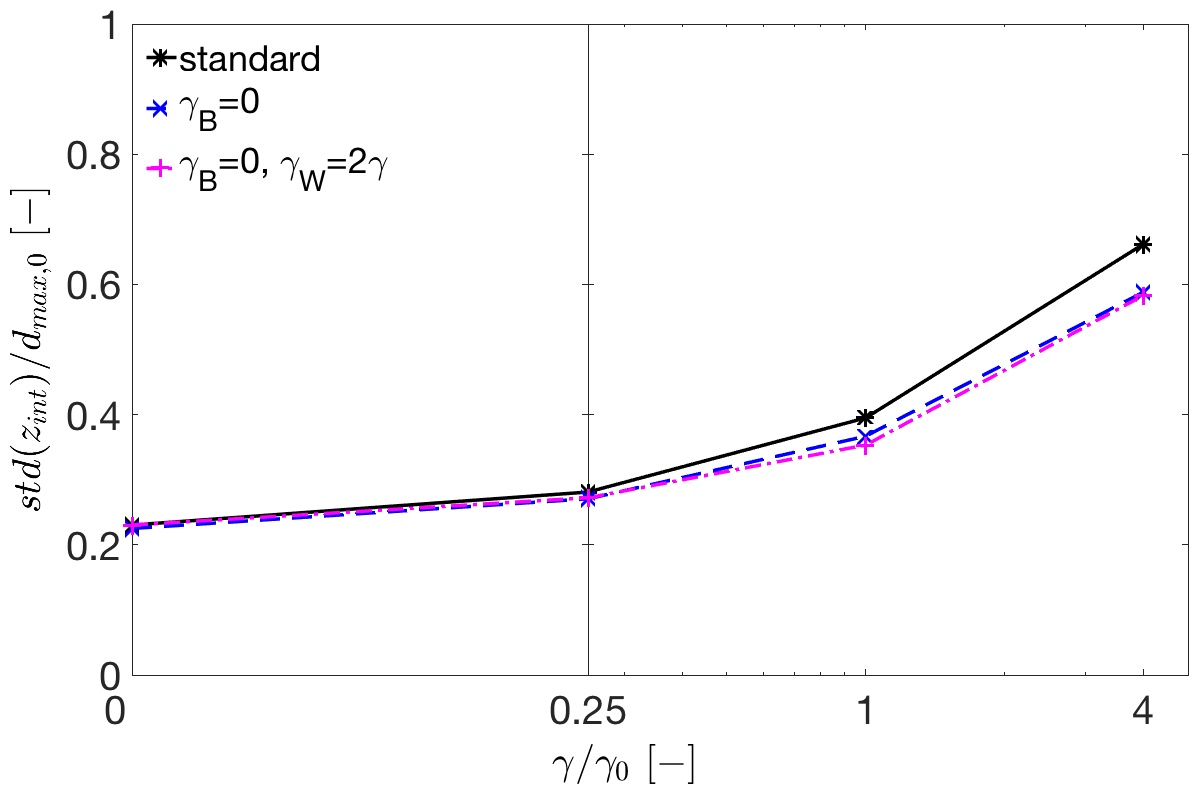}
    \label{fig:wall_4}
   }
   \caption{Mean value and standard deviation of $\Phi_t(x,y)$ and $z_{int}(x,y)$ as function of powder cohesiveness $\gamma / \gamma_0$ for different blade and substrate properties ($t_0=3d_{max,0}$).}
  \label{fig:wall}
\end{figure}

First, the influence of particle-to-blade adhesion shall be investigated. Thereto, the ideal case of a vanishing adhesion between the powder particles and the recoating blade is considered, i.e. $\gamma_B=0$. Here and in the following, a subscript $B$ or $W$ refers to a parameter of particle-to-blade or particle-to-substrate interaction, respectively, while a parameter without subscript refers to particle-to-particle interaction. In Figure~\ref{fig:nba}, the resulting powder layers of the cases $\gamma_B=\gamma$ (left column) and $\gamma_B=0$ (right column) are compared. As expected, the number of surface asperities exceeding the lower edge of the recoating blade is reduced considerably and the surface of the 2D projection of the powder layer in Figure~\ref{fig:nba_2} appears much smoother for the case $\gamma_B=0$ since no powder particles are ripped out of the layer compound anymore. However, Figure~\ref{fig:nba_4} at least indicates that the number and size of depressions in the powder layer is only sightly affected by the particle-to-blade adhesion. The surface profile metrics in Figures~\ref{fig:wall_2} and~\ref{fig:wall_4} confirm this first impression: The mean layer height is slightly increased and the surface roughness is slightly reduced by the choice $\gamma_B=0$. However, in both metrics the level of non-cohesive powders is not reached. While the mean packing fraction seems rarely to be influenced by the choice $\gamma_B=0$, the spatial variation of the packing fraction is reduced noticeably, which might be a consequence of the reduced surface roughness. All in all, the reduction of particle-to-blade adhesion, e.g. by a proper choice of the blade material - seems to be a simple practical measure to improve the powder layer quality for very cohesive / fine-grained powders by a certain degree.\\

In a next step, it shall be investigated if the powder layer quality can be further improved by additionally increasing the adhesion between the powder and the substrate to a value of $\gamma_W=2\gamma$. While the increase of particle-to-substrate adhesion seems to be of rather theoretical nature at first glance, it still allows to answer the question if a potential lack of contact normal forces and / or transferable stick friction might be the bottle neck of the process or not. In practice, comparable effects can be achieved by increasing the powder pile size (yielding higher contact pressure at the interface with the substrate) or the particle-to-substrate friction. According to Figure~\ref{fig:wall}, however, this additional means does not lead to noticeable changes in the considered metrics, or in other words: For the chosen standard parameter set, the particle-to-substrate adhesion already seems to be sufficiently high to avoid excessive relative (slip) motion between the substrate and the contacting particles.\\

\begin{figure}[h!!]
   \centering
   \subfigure[Mean value of packing fraction field $\Phi_t(x,y)$.]
    {
    \includegraphics[width=0.48\textwidth]{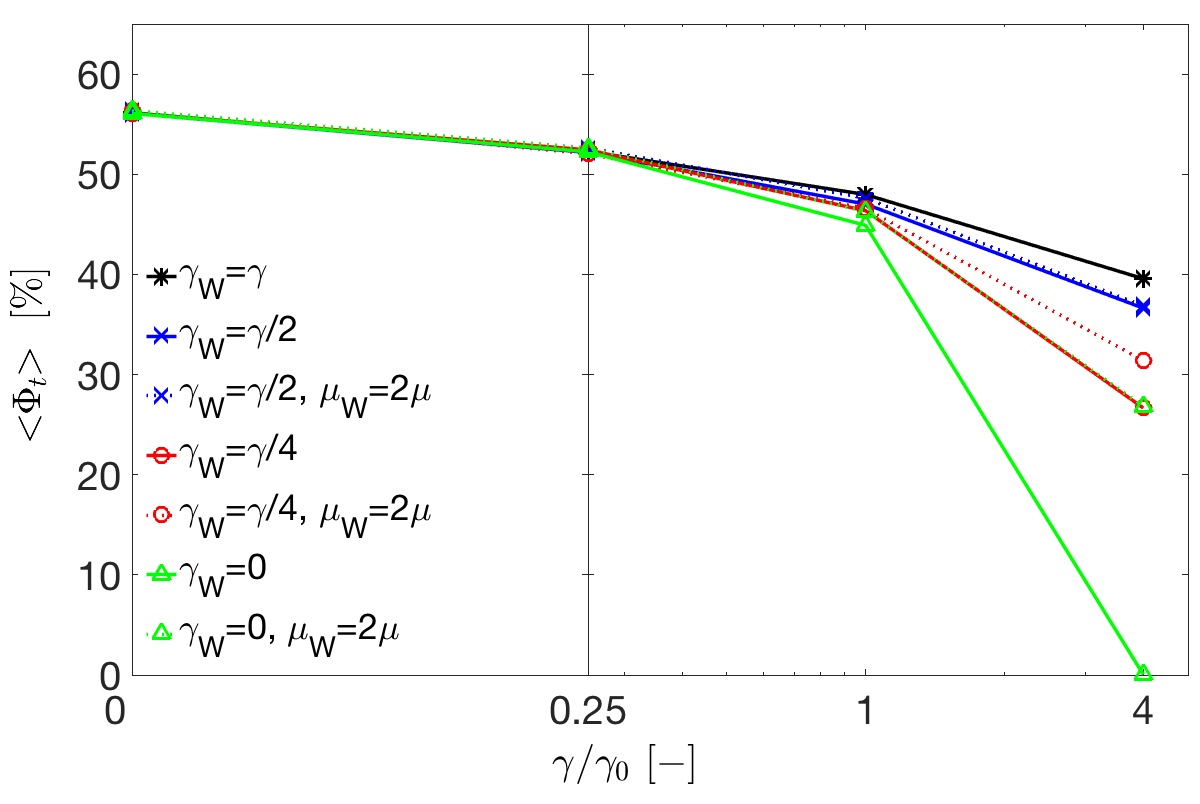}
    \label{fig:wallb_1}
   }
  \hspace{0.1 cm}
   \centering
 \subfigure[Mean value of surface profile field $z_{int}(x,y)$.]
   {
    \includegraphics[width=0.48\textwidth]{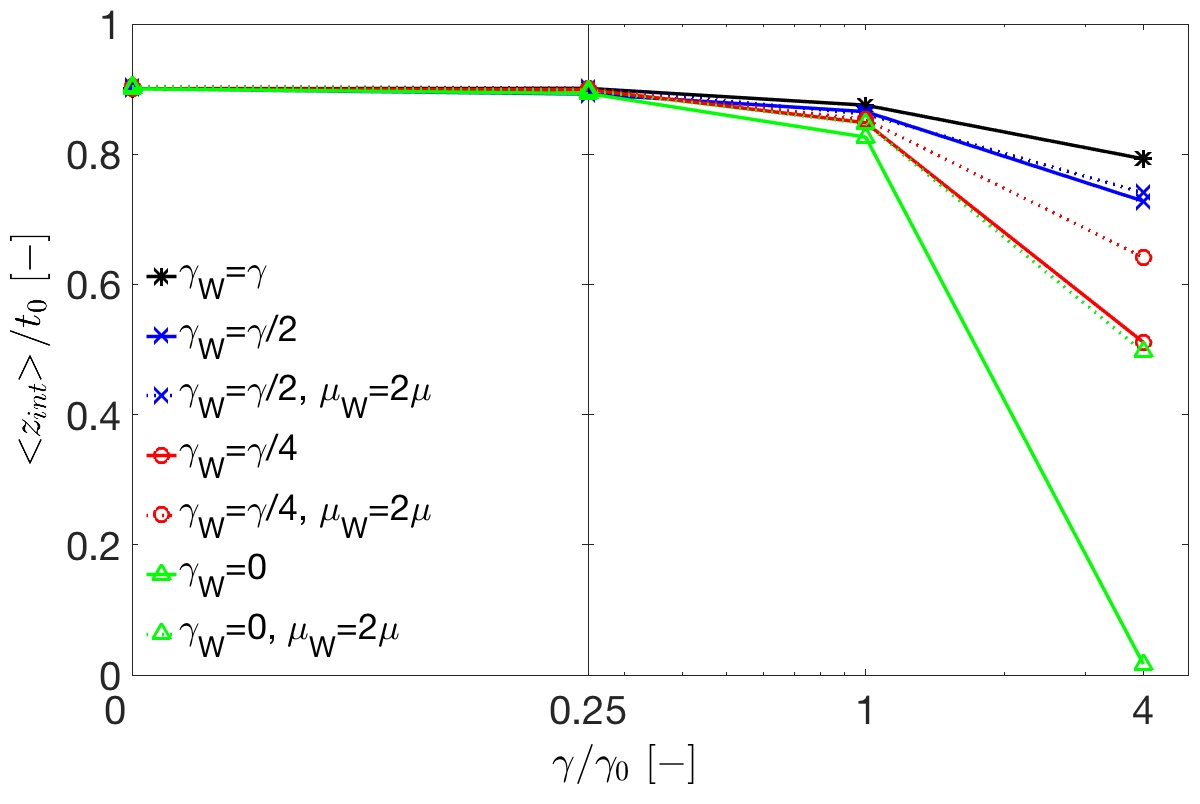}
    \label{fig:wallb_2}
   }    
   \centering
    \subfigure[Standard deviation of packing fraction field $\Phi_t(x,y)$.]
    {
    \includegraphics[width=0.48\textwidth]{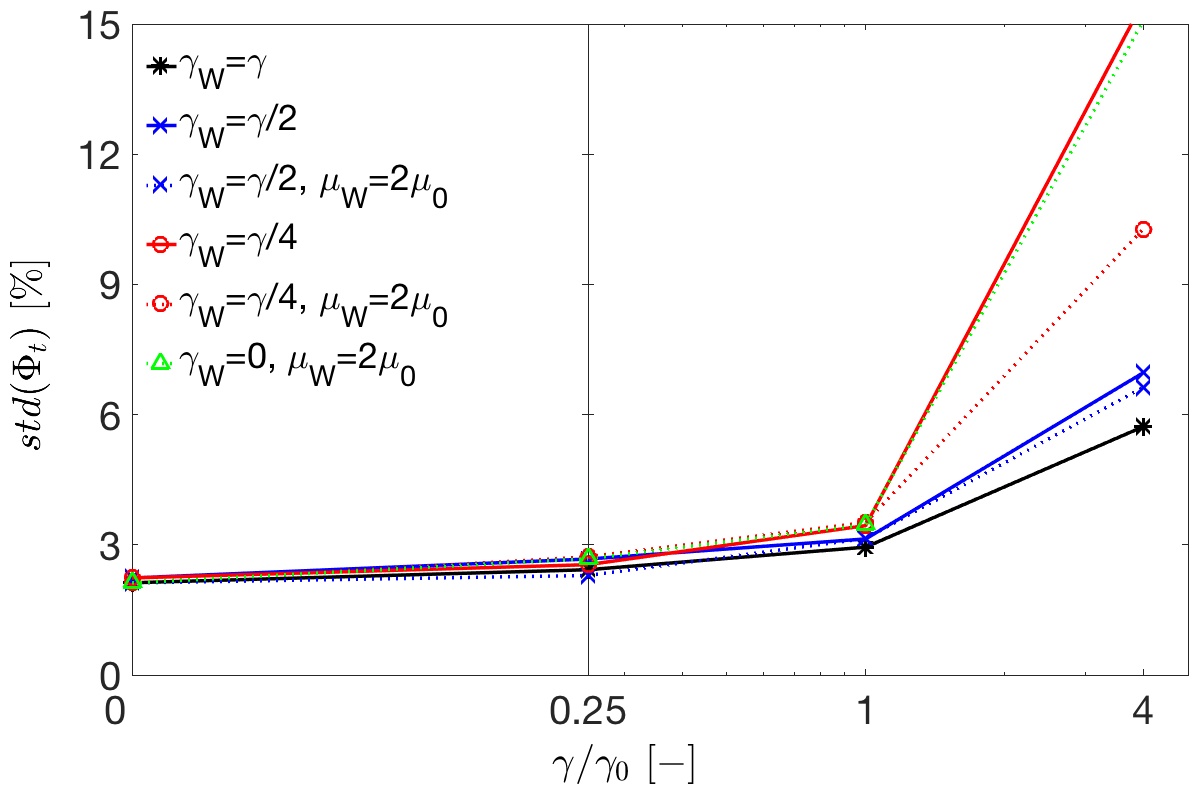}
    \label{fig:wallb_3}
   }
   \hspace{0.1 cm}
 \centering
 \subfigure[Standard deviation of surface profile field $z_{int}(x,y)$.]
   {
    \includegraphics[width=0.48\textwidth]{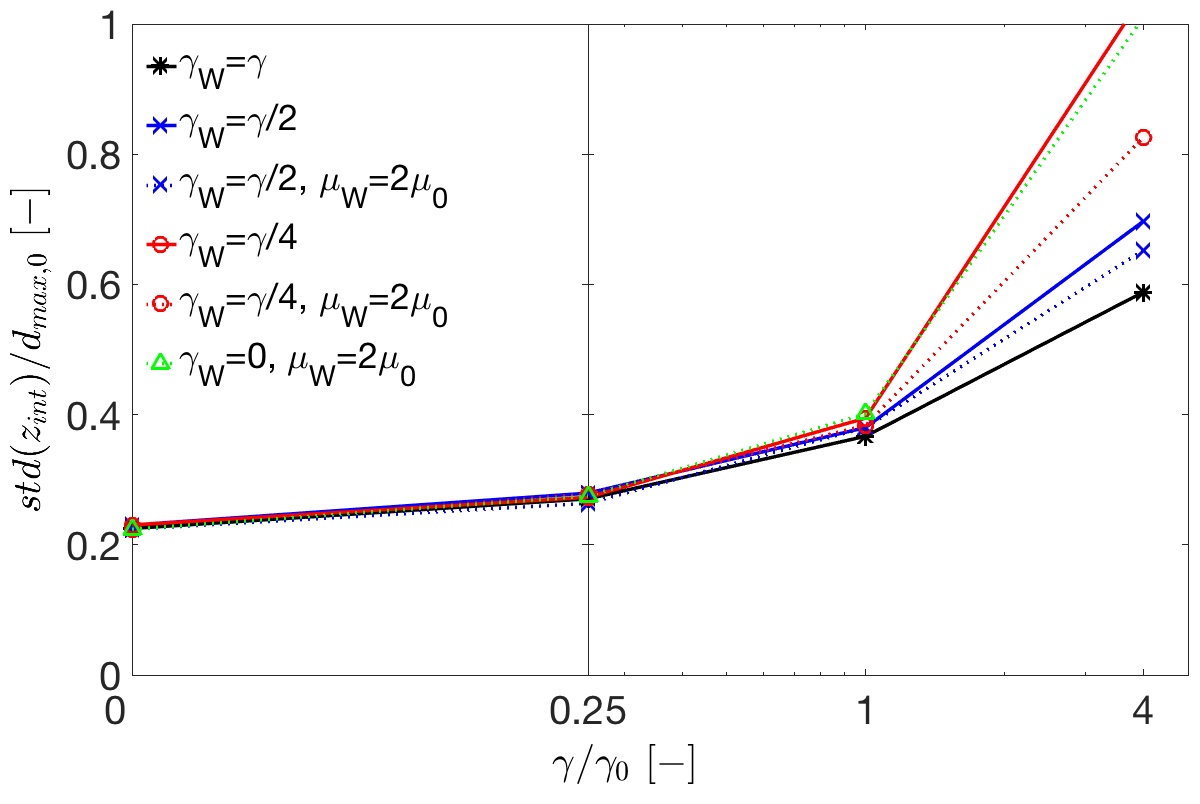}
    \label{fig:wallb_4}
   }
   \caption{Mean value and standard deviation of $\Phi_t(x,y)$ and $z_{int}(x,y)$ as function of powder cohesiveness $\gamma / \gamma_0$ for vanishing blade adhesion ($\gamma_B=0$) and decreased substrate adhesion ($t_0=3d_{max,0}$).}
  \label{fig:wallb}
\end{figure}

In the following investigations, the case of vanishing blade adhesion $\gamma_B=0$ will be considered as reference. In addition to $\gamma_B=0$, now decreased adhesion values (and increased friction values) between the powder and the underlying substrate will be considered. These properties can be designed / optimized for the base plate of the AM machine by means of a proper surface finish, which, however, only influences the recoating process for the very first powder layer. The particle-to-substrate interaction of subsequent layers (here, substrate denotes the solidified cross-section of the design part) rather depends on the parameters of the AM process and is much more complex to control. Typically, higher friction and lower adhesion values due to higher roughness and potential contamination / oxidation of the solidified metal surface can be expected in the latter case. In the following, the particle-to-substrate adhesion values $\gamma_W=\gamma/2$, $\gamma/4$ and $0$ in combination with the friction values $\mu_W=\mu$ and $\mu_W=2\mu$ will be investigated.\\

\begin{figure}[h!!]
   \centering
   \subfigure[Top view: $\gamma_W=\gamma/2, \,\mu_W\!=\!\mu$.]
   {
    \includegraphics[width=0.46\textwidth]{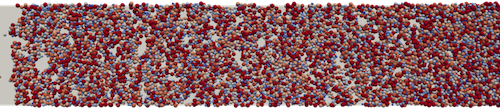}
    \label{fig:nwa_1}
   }
    \hspace{0.1 cm}
    \subfigure[Top view: $\gamma_W=\gamma/4, \,\mu_W\!=\!\mu$.]
    {
    \includegraphics[width=0.46\textwidth]{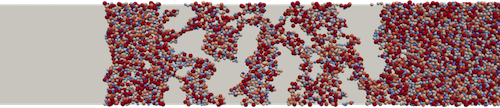}
    \label{fig:nwa_2}
   }
   \caption{Resulting powder layer in dependence of the particle-to-substrate adhesion $\gamma_W$ for the case $\gamma=4\gamma_0$ ($t_0=3d_{max,0}$).}
  \label{fig:nwa}
\end{figure}

In the case $\gamma_W\!=\!0, \, \mu_W\!=\!\mu$ (not illustrated), the powder pile is simply pushed all over the bottom of the powder bed with no powder particles remaining stuck to the underlying substrate. Consequently, the resulting packing fraction and layer height is zero (see Figure~\ref{fig:wallb}). Again, the spatial variations of this case are not representative and will be omitted in the corresponding plots in Figures~\ref{fig:wallb_3} and~\ref{fig:wallb_4}. Increasing the particle-to-substrate adhesion to $\gamma_W\!=\!\gamma/4$ (still at $\mu_W\!=\!\mu$) already improves the level of these metrics noticeable (see again Figure~\ref{fig:wallb}). From Figure~\ref{fig:nwa_1}, it can be seen that the increased adhesion leads to comparatively large areas where the powder sticks to the substrate alternating with approximately equal-sized areas that remain blank. According to Figure~\ref{fig:wallb}, a very similar effect can be achieved by doubling the friction coefficient instead, i.e. $\gamma_W\!=\!0, \,\mu_W\!=\!2\mu$. The adhesion value $\gamma_W\!=\!\gamma/2$ (at $\mu_W\!=\!\mu$) already leads to a powder layer that is very similar to the original one from a qualitative and quantitative point of view (see Figures~\ref{fig:nwa_1} and~\ref{fig:wallb}). In this regime, the influence of increased friction (case $\gamma_W\!=\!\gamma/2,\, \mu_W\!=\!2\mu$) already becomes negligible, i.e. almost no relative (slip) motion between substrate and contacting particles is remaining at this point.\\

All in all, it can be concluded that the recoating of highly cohesive powders is very sensitive with respect to the powder-to-substrate adhesion. Already a decreased adhesion value by a factor of four can lead to a drastically decreased layer quality in terms of lower mean values and higher standard deviations in the packing fraction and surface layer profile. In practical AM processes, potential surface contamination and variations in the surface roughness of the solidified material can easily lead to variations of the powder-to-substrate adhesion by a factor of four and above due to the extreme short-range nature of vdW interactions. It is expected that the consequences of reduced powder-to-substrate adhesion can - at least partly - be compensated by the higher normal forces resulting from larger powder piles sizes and the higher stick friction limit resulting from a higher surface roughness of the solidified material. For less cohesive powders ($\gamma \leq \gamma_0$, Figure~\ref{fig:wallb}), the sensitivity concerning particle-to-substrate adhesion decreases significantly.

\section{Discussion of results}
\label{sec:discussion}

In this section, the main findings of the previously presented recoating simulations will be summarized, and their main implications on the practical AM process will be discussed.\\

For the considered medium-sized powder with mean diameter $\bar{d}=d_0=34\mu m$, the effect of adhesion already plays an important role and has a considerable influence on the resulting powder layer metrics compared to the (theoretical) case, where adhesion has been switched off in the numerical simulations. Moreover, it has been shown that the sensitivity of resulting layer characteristics with respect to practically relevant variations / uncertainties of other powder material parameters, such as friction coefficient, coefficient of restitution or chosen penalty / stiffness parameter, are small compared to the sensitivity with respect to practically relevant variations / uncertainties in the surface energy. Consequently, the present study suggests that the consistent consideration of powder cohesion is vital for realistic recoating simulations, which is in agreement with the results derived by~\cite{Meier2018a}.\\

Taking the powder layers resulting from this medium-sized powder with calibrated surface energy $\gamma=\gamma_0$ as reference, a more fine-grained powder with half the mean particle diameter $\bar{d}=d_0/2$ (or equivalently with $\bar{d}=d_0$ and $\gamma=4\gamma_0$) already leads to a drastically reduced powder layer quality. The decreased mean packing fraction is expected to lead to a lower laser energy absorption~\citep{Boley2015}, i.e. to a less energy-efficient process, as well as to a lower effective thermal conductivity of the powder layer. The latter property might in turn result in higher temperature gradients, increased residual stresses and a reduced conductive energy and temperature homogenization, potentially resulting in local over-/underheating accompanied by unmolten particles / inclusions and evaporated material. In combination with a lower mean surface profile height, the lower packing fraction leads to increased shrinkage of the powder layer during the melting (and solidification) process. While the reduced mean values in packing fraction and surface profile height can - at least partly - be compensated by adapted machine parameters (e.g. laser beam power, velocity etc.), the increased standard deviations in these metrics might lead to considerable spatial variations in the resulting layer height of the solidified material but also to strong variations in effective laser energy absorptivity and thermal conductivity, again fostering local over-/underheating as described above. On the one hand, high layer thickness variations and large surface asperities as visible e.g. in Figure~\ref{fig:sv_8} require a high laser energy density to achieve complete powder melting and sufficient remelting of the substrate even at locations of high layer thickness. On the other hand, single powder particles at the top of surface asperities at such locations are practically thermally insulated from the rest of the powder layer and consequently prone to (undesirable) evaporation. All in all, it can be concluded that the powder layers resulting from the spreading of such fine / cohesive powders on the basis of a simple recoating blade without additional means of powder compaction are not suitable for subsequent laser melting.\\

This argumentation applies even more when realizing that most of the studies have been conducted on the assumption of an ideal powder-to-substrate interaction. As shown in Section~\ref{sec:recoating_bladeandsubstrate}, a reduction of powder-to-substrate adhesion by a factor of four, which can easily occur due to contamination / oxidation on the surface of the solidified material in the practically relevant range, already leads to strongly discontinuous powder layers with considerable blank spots. It has been shown that this effect can (at least partly) be compensated by an increased level of particle-to-substrate friction. This also means that the different effects of increased surface roughness - namely lower adhesion and higher friction - might compensate each other up to a certain degree. For the investigated medium-sized and coarse-grained powder, the observed sensitivity with respect to powder-to-substrate adhesion was much lower. For these powders, powder-to-substrate friction is expected to play a more important role than powder-to-substrate adhesion.\\

On the other hand, it has been shown that the layer quality of very fine-grained / cohesive powders can be improved up to a certain degree by reducing the adhesion between powder particles and the recoating blade. This can e.g. be realized by choosing a proper material for the recoating blade. From Figure~\ref{fig:nba}, it becomes obvious that at least the effect of very strong layer thickness variations and isolated particles at the top of surface asperities, as discussed above, might be improved by this means. In combination with a subsequent powder compaction step, this measure might considerably improve the spreadability and resulting layer characteristics of very fine / cohesive powders. Again, the influence of this measure on rather coarse-grained / less-adhesive powders is limited.\\

The investigation of different layer thicknesses revealed that very low layer thicknesses $t_0=d_{max,0}$, only slightly above the maximal particle diameter of the considered size distribution, lead to very sparse, discontinuous powder layers, which are not suitable for subsequent laser melting. This observation was independent of the considered powder size / cohesion level. For medium-sized and coarse-grained powders, nominal layer thicknesses $t_0=2d_{max,0}$ in the range of twice the maximal particle diameter and above already lead to continuous powder layers of good quality. By further increasing the layer thickness, the quality (e.g. in terms of packing fraction and mean layer height) still increases slightly due to decreased boundary effects on the top and bottom of the powder layer. However, optimal layer characteristics are already approached very closely at layer thickness of two to three times the maximal powder particle diameter. For the investigated fine-grained powder, the recommended minimal layer thickness rather lies in the range of three to four times the maximal particle diameter. Moreover, also the sensitivity e.g. of the resulting packing fraction field with respect to (too) low powder layer thicknesses has been observed to increase with increasing cohesiveness of the powder.\\

Eventually, also the effect of blade velocity has been investigated in the present study. It was found that an increased blade velocity leads to a dynamic post-flow of (already spread) powder, which, in turn, results in a powder layer of lower mean layer thickness. Accordingly, the remaining powder metrics show similar trends as for the case of reduced layer thicknesses discussed above. Further, it was found that this undesirable effect decreases with decreasing flowability of the powder. It was shown that flowability can be reduced and layer characteristics can be (slightly) improved in the range of high blade velocities by increasing inter-particle friction. Stronger influence on flowability could e.g. be exerted by applying powders with less spherical particles, requiring a compromise between controlled / reduced flowability and achievable packing fractions (which typically decrease for less spherical particles). Of course, the effect of powder post-flow and reduced layer thicknesses in the high velocity regime could also be compensated by applying correspondingly increased nominal layer thicknesses / blade gaps. Eventually, it was found that this effect was less pronounced in the range of very fine-grained / cohesive powders, i.e. the cohesiveness of powder can also provide potential benefits for the recoating process.\\

\section{Conclusion}
\label{sec:conclusion}

The present work employed a cohesive powder model recently proposed by the authors to study the powder recoating process in powder-bed fusion metal additive manufacturing (AM), and to analyze the influence of different process parameters with a special focus on cohesion. The employed model is based on the discrete element method (DEM) with particle-to-particle and particle-to-wall interactions involving frictional contact, rolling resistance and adhesive forces. While most model parameters are not particularly sensitive and hence have been taken from the literature, the surface energy defining the adhesive interaction forces between particles can easily vary by several orders of magnitude for a given material combination. Therefore, the value of the effective surface energy has been taken from the authors' previous work~\citep{Meier2018a}, where a model calibration has been conducted for the considered powder material by fitting angle of repose (AOR) values from numerical and experimental funnel tests.\\

Building on this pre-work, the present contribution represents the first numerical study of powder recoating processes in metal additive manufacturing based on a realistic powder model that considers the important effect of powder cohesiveness. Besides the calibrated surface energy value $\gamma=\gamma_0$ and the theoretical reference scenario $\gamma=0$, also the cases of decreased / increased surface energies $\gamma=\gamma_0/4$ and $\gamma=4\gamma_0$ have been considered in the recoating simulations. These cases are equivalent to increasing / decreasing the mean powder particle diameter from $\bar{d}=\bar{d}_0=34\mu m$ to $\bar{d}=2\bar{d}_0$ and $\bar{d}=\bar{d}_0/2$, respectively. In order to evaluate the quality of the resulting powder layers, proper metric definitions in terms of the spatial surface profile and packing fraction field have been proposed. Besides particle-to-particle adhesion, the influence of mechanical bulk powder material parameters (friction coefficient, coefficient of restitution, particle stiffness parameter), the nominal layer thickness, the blade velocity as well as particle-to-wall adhesion / friction has been analyzed in comprehensive parameter studies. The following main results have been derived in this contribution:

\begin{itemize}
\item Increased cohesiveness / decreased particle size leads to considerably decreased powder layer quality in terms of low, strongly varying packing fractions and high surface roughness / non-uniformity. Resulting packing fractions are even below the values typically known for (unspread) bulk powder of comparable cohesiveness.
\item The cohesive forces in the most fine-grained powder with mean diameter $\bar{d}=\bar{d}_0/2=17\mu m$ dominate gravity forces by almost two orders of magnitude. The resulting powder layers are of very low quality and seem not to be suitable for subsequent laser melting without additional layer / surface finishing steps.
\item The powder layer quality further decreases if the adhesion between powder and substrate is impaired e.g due to surface contamination / oxidation.
\item The surface uniformity of cohesive powder layers can be improved by designing recoating blades with reduced adhesive interaction.
\item The powder layer quality increases with increasing nominal layer thickness. A minimal nominal layer thickness of two to three times the maximal powder particle diameter is recommended for optimal layer quality.
\item Increased blade velocities lead to dynamic powder post-flow and decreased mean layer thicknesses, which can be compensated by reduced powder flowability or increased nominal layer thickness.
\end{itemize}

All in all, the present study has given valuable insights into the powder recoating process in metal AM. The role of cohesion and other material and process parameters has been analyzed, and practical implications for the subsequent melting process have been derived. Future research work will focus on means of experimental verification and on the analysis and optimization of alternative recoating tools and strategies.

\section*{Acknowledgements}
\label{sec:Acknowledgements}

Financial support for preparation of this article was provided by the German Academic Exchange Service (DAAD) to C. Meier and R. Weissbach as well as by Honeywell Federal Manufacturing \& Technologies to A.J. Hart. This work has been partially funded by Honeywell Federal Manufacturing \& Technologies, LLC which manages and operates the Department of Energy's Kansas City National Security Campus under Contract No. DE-NA-0002839. The United States Government retains and the publisher, by accepting the article for publication, acknowledges that the United States Government retains a nonexclusive, paid up, irrevocable, world-wide license to publish or reproduce the published form of this manuscript, or allow others to do so, for the United States Government purposes. This material is based upon work supported by the Assistant Secretary of Defense for Research and Engineering under Air Force Contract No. FA8721-05-C-0002 and/or FA8702-15-D-0001. Any opinions, findings, conclusions or recommendations expressed in this material are those of the author(s) and do not necessarily reflect the views of the Assistant Secretary of Defense for Research and Engineering.


%
\bibliography{powdercoating_1.bib,review_heat_transfer_Jan2018_2.bib,beamreferences.bib}

\begin{thebibliography}{45}
\providecommand{\natexlab}[1]{#1}

\bibitem[{Boley et~al.(2015)Boley, Khairallah, and Rubenchik}]{Boley2015}
Boley, C., Khairallah, S., Rubenchik, A. (2015). {Calculation of laser
  absorption by metal powders in additive manufacturing.}, Applied optics 54,
  2477--2482.

\bibitem[{Cundall and Strack(1979)}]{Cundall1979}
Cundall, P., Strack, O. (1979). {A discrete numerical model for granular
  assemblies}, G{\'{e}}otechnique 29, 47--65.

\bibitem[{Das(2003)}]{Das2003}
Das, S. (2003). {Physical Aspects of Process Control in Selective Laser
  Sintering of Metals}, Advanced Engineering Materials 5, 701--711.

\bibitem[{Denlinger et~al.(2015)Denlinger, Heigel, and
  Michaleris}]{Denlinger2014}
Denlinger, E., Heigel, J., Michaleris, P. (2015). {Residual stress and
  distortion modeling of electron beam direct manufacturing Ti-6Al-4V},
  Proceedings of the Institution of Mechanical Engineers, Part B: Journal of
  Engineering Manufacture 229, 1803--1813.

\bibitem[{Ebert et~al.(2003)Ebert, Regenfuss, Kloetzer, Hartwig, and
  Exner}]{Ebert2003}
Ebert, R., Regenfuss, P., Kloetzer, S., Hartwig, L., Exner, H. (2003). {Process
  assembly for $\mu$m-scale SLS, reaction sintering, and CVD}, in Fourth
  International Symposium on Laser Precision Microfabrication, pp. 183--189.

\bibitem[{Gibson et~al.(2010)Gibson, Rosen, and Stucker}]{Gibson2010}
Gibson, I., Rosen, D., Stucker, B. (2010). {Additive manufacturing
  technologies}, volume 238, Springer.

\bibitem[{Gong and Chou(2015)}]{Gong2015}
Gong, X., Chou, K. (2015). {Phase-field modeling of microstructure evolution in
  electron beam additive manufacturing}, JOM 67, 1176--1182.

\bibitem[{Gunasegaram et~al.(2017)Gunasegaram, Murphy, Cummins, Lemiale,
  Delaney, Nguyen, and Feng}]{Gunasegaram2017}
Gunasegaram, D., Murphy, A., Cummins, S., Lemiale, V., Delaney, G., Nguyen, V.,
  Feng, Y. (2017). {Aiming for Modeling-Assisted Tailored Designs for Additive
  Manufacturing}, in M.{\&}.M.S. {TMS The Minerals}, editor, TMS 2017 146th
  Annual Meeting {\&} Exhibition Supplemental Proceedings, pp. 91--102, Cham.

\bibitem[{Gusarov(2008)}]{Gusarov2008}
Gusarov, A. (2008). {Homogenization of radiation transfer in two-phase media
  with irregular phase boundaries}, Physical Review B - Condensed Matter and
  Materials Physics 77, 1--14.

\bibitem[{Gusarov and Kruth(2005)}]{Gusarov2005}
Gusarov, A., Kruth, J. (2005). {Modelling of radiation transfer in metallic
  powders at laser treatment}, International Journal of Heat and Mass Transfer
  48, 3423--3434.

\bibitem[{Gusarov et~al.(2003)Gusarov, Laoui, Froyen, and Titov}]{Gusarov2003}
Gusarov, A., Laoui, T., Froyen, L., Titov, V. (2003). {Contact thermal
  conductivity of a powder bed in selective laser sintering}, International
  Journal of Heat and Mass Transfer 46, 1103--1109.

\bibitem[{Gusarov et~al.(2007)Gusarov, Yadroitsev, Bertrand, and
  Smurov}]{Gusarov2007}
Gusarov, A., Yadroitsev, I., Bertrand, P., Smurov, I. (2007). {Heat transfer
  modelling and stability analysis of selective laser melting}, Applied Surface
  Science 254, 975--979.

\bibitem[{Haeri(2017)}]{Haeri2017}
Haeri, S. (2017). {Optimisation of blade type spreaders for powder bed
  preparation in Additive Manufacturing using DEM simulations}, Powder
  Technology 321, 94--104.

\bibitem[{Herbert(2016)}]{Herbert2016}
Herbert, R. (2016). {Viewpoint: metallurgical aspects of powder bed metal
  additive manufacturing}, Journal of Materials Science 51, 1165--1175.

\bibitem[{Herbold et~al.(2015)Herbold, Walton, and Homel}]{Herbold2015}
Herbold, E., Walton, O., Homel, M. (2015). {Simulation of Powder Layer
  Deposition in Additive Manufacturing Processes Using the Discrete Element
  Method}, Technical report, Lawrence Livermore National Laboratory.

\bibitem[{Hodge et~al.(2014)Hodge, Ferencz, and Solberg}]{Hodge2014}
Hodge, N., Ferencz, R., Solberg, J. (2014). {Implementation of a
  thermomechanical model for the simulation of selective laser melting},
  Computational Mechanics 54, 33--51.

\bibitem[{Israelachvili(2011)}]{Israelachvili2011}
Israelachvili, J. (2011). {Intermolecular and Surface Forces}, Academic Press,
  San Diego, 3 edition.

\bibitem[{Khairallah and Anderson(2014)}]{Khairallah2014}
Khairallah, S., Anderson, A. (2014). {Mesoscopic simulation model of selective
  laser melting of stainless steel powder}, Journal of Materials Processing
  Technology 214, 2627--2636.

\bibitem[{Khairallah et~al.(2016)Khairallah, Anderson, Rubenchik, and
  King}]{Khairallah2016}
Khairallah, S., Anderson, A., Rubenchik, A., King, W. (2016). {Laser powder-bed
  fusion additive manufacturing: Physics of complex melt flow and formation
  mechanisms of pores, spatter, and denudation zones}, Acta Materialia 108,
  36--45.

\bibitem[{King(2017)}]{King2017}
King, W. (2017). {Modeling of Powder Dynamics in Metal Additive Manufacturing:
  Final Powder Dynamics Meeting Report,
  https://hpc4mfg.llnl.gov/events-powderdynamics2017.php}.

\bibitem[{K{\"{o}}rner et~al.(2011)K{\"{o}}rner, Attar, and Heinl}]{Korner2011}
K{\"{o}}rner, C., Attar, E., Heinl, P. (2011). {Mesoscopic simulation of
  selective beam melting processes}, Journal of Materials Processing Technology
  211, 978--987.

\bibitem[{Kruth et~al.(2007)Kruth, Levy, Klocke, and Childs}]{Kruth2007}
Kruth, J., Levy, G., Klocke, F., Childs, T. (2007). {Consolidation phenomena in
  laser and powder-bed based layered manufacturing}, CIRP Annals -
  Manufacturing Technology 56, 730--759.

\bibitem[{Lechman et~al.(2013)Lechman, Yarrington, Erikson, and
  Noble}]{Lechman2013}
Lechman, J., Yarrington, C., Erikson, W., Noble, D. (2013). {Thermal conduction
  in particle packs via finite elements}, AIP Conference Proceedings 1542,
  539--542.

\bibitem[{Lee(2015)}]{Lee2015}
Lee, Y. (2015). {Simulation of Laser Additive Manufacturing and its
  Applications}, Ph.D. thesis, The Ohio State University.

\bibitem[{Matthews et~al.(2016)Matthews, Guss, Khairallah, Rubenchik, Depond,
  and King}]{Matthews2016}
Matthews, M., Guss, G., Khairallah, S., Rubenchik, A., Depond, P., King, W.
  (2016). {Denudation of metal powder layers in laser powder bed fusion
  processes}, Acta Materialia 114, 33--42.

\bibitem[{Meier et~al.(2018{\natexlab{a}})Meier, Penny, Zou, Gibbs, and
  Hart}]{Meier2018}
Meier, C., Penny, R., Zou, Y., Gibbs, J., Hart, A. (2018{\natexlab{a}}).
  {Thermophysical Phenomena in Metal Additive Manufacturing by Selective Laser
  Melting: Fundamentals, Modeling, Simulation and Experimentation}, Annual
  Review of Heat Transfer, DOI: 10.1615/AnnualRevHeatTransfer.2018019042 .

\bibitem[{Meier et~al.(2018{\natexlab{b}})Meier, Weissbach, Weinberg, Wall, and
  Hart}]{Meier2018a}
Meier, C., Weissbach, R., Weinberg, J., Wall, W., Hart, A.
  (2018{\natexlab{b}}). {Modeling and Characterization of Cohesion in Fine
  Metal Powders with a Focus on Additive Manufacturing Process Simulations},
  submitted to Powder Technology, arXiv:1804.06816 .

\bibitem[{Meyer et~al.(2017)Meyer, Wegner, and Witt}]{Meyer2017}
Meyer, L., Wegner, A., Witt, G. (2017). {Influence of the Ratio between the
  Translation and Contra-Rotating Coating Mechanism on different Laser
  Sintering Materials and their Packing Density}, in Solid Freeform Fabrication
  2017: Proceedings of the 28th Annual International Solid Freeform Fabrication
  Symposium – An Additive Manufacturing Conference, pp. 1432--1447, Austin.

\bibitem[{Mindt et~al.(2016)Mindt, Megahed, Lavery, Holmes, and
  Brown}]{Mindt2016}
Mindt, H., Megahed, M., Lavery, N., Holmes, M., Brown, S. (2016). {Powder Bed
  Layer Characteristics: The Overseen First-Order Process Input}, Metallurgical
  and Materials Transactions A: Physical Metallurgy and Materials Science 47,
  3811--3822.

\bibitem[{Neef et~al.(2014)Neef, Seyda, Herzog, Emmelmann, Sch{\"{o}}nleber,
  and Kogel-Hollacher}]{Neef2014}
Neef, A., Seyda, V., Herzog, D., Emmelmann, C., Sch{\"{o}}nleber, M.,
  Kogel-Hollacher, M. (2014). {Low coherence interferometry in selective laser
  melting}, Physics Procedia 56, 82--89.

\bibitem[{O'Sullivan and Bray(2004)}]{OSullivan2004}
O'Sullivan, C., Bray, J. (2004). {Selecting a suitable time step for discrete
  element simulations that use the central difference time integration
  scheme.}, Engineering Computations 21, 278--303.

\bibitem[{Parteli and P{\"{o}}schel(2016)}]{Parteli2016}
Parteli, E., P{\"{o}}schel, T. (2016). {Particle-based simulation of powder
  application in additive manufacturing}, Powder Technology 288, 96--102.

\bibitem[{Qiu et~al.(2015)Qiu, Panwisawas, Ward, Basoalto, Brooks, and
  Attallah}]{Qiu2015}
Qiu, C., Panwisawas, C., Ward, M., Basoalto, H., Brooks, J., Attallah, M.
  (2015). {On the role of melt flow into the surface structure and porosity
  development during selective laser melting}, Acta Materialia 96, 72--79.

\bibitem[{Rai et~al.(2016)Rai, Markl, and K{\"{o}}rner}]{Rai2016}
Rai, A., Markl, M., K{\"{o}}rner, C. (2016). {A coupled Cellular
  Automaton-Lattice Boltzmann model for grain structure simulation during
  additive manufacturing}, Computational Materials Science 124, 37--48.

\bibitem[{Riedlbauer et~al.(2016)Riedlbauer, Scharowsky, Singer, Steinmann,
  K{\"{o}}rner, and Mergheim}]{Riedlbauer2016}
Riedlbauer, D., Scharowsky, T., Singer, R., Steinmann, P., K{\"{o}}rner, C.,
  Mergheim, J. (2016). {Macroscopic simulation and experimental measurement of
  melt pool characteristics in selective electron beam melting of Ti-6Al-4V},
  International Journal of Advanced Manufacturing Technology pp. 1--9.

\bibitem[{Tan et~al.(2017)Tan, Wong, and Dalgarno}]{Tan2017}
Tan, J., Wong, W., Dalgarno, K. (2017). {An overview of powder granulometry on
  feedstock and part performance in the selective laser melting process},
  Additive Manufacturing 18, 228--255.

\bibitem[{Thomas et~al.(2016)Thomas, Baxter, and Todd}]{Thomas2016}
Thomas, M., Baxter, G., Todd, I. (2016). {Normalised model-based processing
  diagrams for additive layer manufacture of engineering alloys}, Acta
  Materialia 108, 26--35.

\bibitem[{Wall(2018)}]{Wall2018}
Wall, W. (2018). {BACI: A multiphysics simulation environment.}, Technical
  report, Technical University of Munich.

\bibitem[{Walton(2008)}]{Walton2008}
Walton, O. (2008). {Review of Adhesion Fundamentals for Micron Scale
  Particles}, Kona Powder and Particle Journal 26, 129--141.

\bibitem[{Walton et~al.(2007)Walton, {De Moor}, and Gill}]{Walton2007}
Walton, O., {De Moor}, C., Gill, K. (2007). {Effects of gravity on cohesive
  behavior of fine powders: Implications for processing Lunar regolith},
  Granular Matter 9, 353--363.

\bibitem[{Whiting and Fox(2016)}]{Whiting2016}
Whiting, J., Fox, J. (2016). {Characterization of Feedstock in the Powder Bed
  Fusion Process: Sources of Variation in Particle Size Distribution and the
  Factors that Influence them}, in International Solid Freeform Fabrication
  Symposium, Austin, Texas, USA.

\bibitem[{Yablokova et~al.(2015)Yablokova, Speirs, {Van Humbeeck}, Kruth,
  Schrooten, Cloots, Boschini, Lumay, and Luyten}]{Yablokova2015}
Yablokova, G., Speirs, M., {Van Humbeeck}, J., Kruth, J., Schrooten, J.,
  Cloots, R., Boschini, F., Lumay, G., Luyten, J. (2015). {Rheological behavior
  of $\beta$-Ti and NiTi powders produced by atomization for SLM production of
  open porous orthopedic implants}, Powder Technology 283, 199--209.

\bibitem[{Yang et~al.(2003)Yang, Zou, and Yu}]{Yang2000}
Yang, R., Zou, R., Yu, A. (2003). {Computer simulation of the packing of
  particles}, International Journal of Materials and Product Technology 19,
  324.

\bibitem[{Zaeh and Branner(2010)}]{Zaeh2010}
Zaeh, M., Branner, G. (2010). {Investigations on residual stresses and
  deformations in selective laser melting}, Production Engineering 4, 35--45.

\bibitem[{Zhirnov et~al.(2018)Zhirnov, Kotoban, and Gusarov}]{Zhirnov2018}
Zhirnov, I., Kotoban, D., Gusarov, A. (2018). {Evaporation-induced gas-phase
  flows at selective laser melting}, Applied Physics A: Materials Science and
  Processing 124, 157.

\end{thebibliography}
%
%
\end{document}